\documentclass[prd,superscriptaddress]{revtex4}
\usepackage{amssymb,amsmath}
\usepackage{graphicx}
\usepackage{color}
\begin{document}

\title{Beryllium isotopic composition \\
 and Galactic cosmic ray propagation}

\author{Paolo Lipari}
\email{paolo.lipari@roma1.infn.it}
\affiliation{INFN sezione Roma ``Sapienza''} 

\begin{abstract}
The isotopic composition of beryllium nuclei and its energy
dependence encode information of fundamental importance
about the propagation of cosmic rays in the Galaxy.
The effects of decay on the spectrum of the unstable
beryllium--10  isotope can be described introducing
the average survival probability
$P_{\rm surv} (E_0)$ that can inferred from
measurements of the isotopic ratio Be10/Be9 if one has 
sufficiently  good knowledge of the
nuclear fragmentation cross sections that determine
the isotopic composition of beryllium nuclei at injection.
The average survival probability can then be interpreted in terms of
propagation parameters, such as the cosmic ray average age, 
adopting a theoretical framework for Galactic propagation.
Recently the AMS02 Collaboration has presented preliminary measurements
of the beryllium isotopic composition that extend the observations
to a broad energy range ($E_0 \simeq 0.7$--12~GeV/n) with
small errors. In this work we discuss the average survival
probability that can be inferred from the preliminary AMS02 data,
adopting publically available models of the nuclear fragmentation
cross sections,
and interpret the results in the framework of a  simple diffusion model, 
This study shows  that the effects of decay decrease more slowly
than the predictions, resulting in an average  cosmic ray age
that increases with energy.
An alternative possibility
is that the cosmic ray age distribution
is broader than in the models that are now commonly accepted,
suggesting that the Galactic confinement volume has a non trivial
structure and is formed by an inner halo contained in an extended one.
\end{abstract}

\maketitle

\section{Introduction}
\label{sec:introduction}
It is now well established that most of the cosmic rays (CR)
observed at the Earth in a broad energy range that
extends from $E \sim 10^9$~eV to at least $E \sim 10^{16}$~eV
are of Galactic origin, and are generated in the Milky Way, where they remain
partially confined by interstellar magnetic fields
for a time of order 1--100~Myr.
Understanding the properties of CR propagation, and determining
the duration and energy (or rigidity) dependence
of their Galactic residence time remains a 
problem of crucial importance for high energy astrophysics.

The study of the flux of the unstable nucleus beryllium--10 (Be10)
has been recognised for a long time as a crucially important
source of information about the properties of CR propagation.
This is because the Be10 decay time ($T_{1/2} \simeq 1.387 \pm 0.012$~Myr)
is comparable with the average CR Galactic residence time, and 
therefore decay can be a significant, or dominant ``sink'' mechanism
in the formation of the spectrum. 
Comparing the spectral shape of Be10 with those of the stable isotopes Be9 and Be7,
allows in principle to measure the effects of decay, and then
infer properties of Galactic propagation.

The experimental study of the spectra of individual isotopes,
is however a very difficult task, and until now measurements
for beryllium have been obtained only at low energy
(kinetic energy per nucleon $E_0 \lesssim 2$~GeV)
and with rather large errors.
Recently, at the 37th International Cosmic Ray Conference in Berlin,
the AMS02 Collaboration has presented preliminary measurements of the 
beryllium isotopes spectra and of the Be10/Be9 ratio
with small errors (of order 10--20\%).
and in a broad energy range ($E_0 \simeq 0.7$--12~GeV).
These results can be of great value to find answers to
some important open questions about CR Galactic propagation.

In this work, waiting for the publication of the AMS02 observations on
the isotopically separated beryllium spectra,
we discuss the preliminary results presented at the ICRC, and the
best methods to study their astrophysical implications.

We argue here that it is both convenient and appropriate to
divide this study into two steps.
In the first step, one starts from measurements
of the isotopic ratio Be10/Be9, to estimate 
the average survival probability $P_{\rm surv} (E_0)$,
a quantity that describes the effects of decay on the Be10 spectrum.
The main uncertainty in this first step is associated to
the description of the nuclear fragmentation cross sections
that determine the beryllium isotopic ratio at production.
In the second step one interprets
the results on $P_{\rm surv}$ to estimate
CR propagation parameters.
This second step is model dependent and is possible only
assuming a theoretical framework that must be carefully discussed.

This paper is organised as follows: in the next section we
define the average survival probability $P_{\rm surv} (E_0)$
that encodes the effects of decay of the spectrum of
the unstable beryllium--10 isotope,
and discuss how it is possible to infer $P_{\rm surv}$ from
measurements of the isotopic Be10/Be9.

In section~\ref{sec:interpretation} we discuss the (energy dependent)
cosmic ray age distribution and how it determines
the average survival probability.

The following section discusses in detail
the 1-Dimensional ``Minimal Diffusion Model'' where propagation
(for particles at a fixed energy) is described by two parameters:
a diffusion time $T_{\rm diff}$ for escape from a homogeneous
Galactic confinement volume, and the vertical size $Z_{\rm halo}$ of
this volume. This model captures the main features of the
models that are in common use to interpret cosmic ray measurements,
but is also sufficiently simple that it is possible to
calculate the average survival probability (and several
other interesting quantities) obtaining exact analytic expressions.
This can be both convenient and instructive, to develop an
understanding of the problem.

In section \ref{sec:ams} we use these results to compute
allowed intervals for the diffusion time and the halo size
that can be inferred from the AMS02 preliminary data.
The main source of systematic error in this exercise is the
estimate of the nuclear fragmentation cross sections that are used
to obtain the average survival probability from the measurements
of the isotopic ratio.

The final section discusses critically the results, and their possible
implications. The most intriguing result that emerges from
the preliminary AMS02 measurements is that the isotopic ratio Be9/Be10
grows with energy more slowly than expectations 
based on current diffusion based models.
If these results are confirmed, this discrepancy
can perhaps be explained as the effect of an incorrect
description of the nuclear fragmentation cross sections.
The alternative possibility is that 
the diffusion models commonly used too interpret the CR observations
are not adequate and must be revised.

\section{From the isotopic ratio to the average survival probability} 
\label{sec:psurv}
In this paper we argue that it is natural
and convenient to study the effects of decay on the Be10 spectrum,
introducing (following \cite{Lipari:2014zna}) the
average survival probability $P_{\rm surv} (E_0)$
(with $E_0$ the kinetic energy per nucleon).
This quantity is defined as:
\begin{equation}
 P_{\rm surv} (E_0) =
\frac{\phi_{10} (E_0)} {\phi_{10}^{(0)} (E_0)} 
\label{eq:psurv-def}
\end{equation}
where the numerator is the Be10 flux at the boundary of the heliosphere
after correcting for solar modulation effects, and the
denominator is the same flux {\em calculated} under the hypothesis that
the nuclei are stable.

This definition might appear problematic, because the denominator
in Eq.~(\ref{eq:psurv-def}) is not a directly measurable quantity,
however this difficulty can be circumvented,
estimating the ``no-decay'' Be10 flux, using the observed
flux of the stable isotope beryllium--9 and applying appropriate corrections,
as discussed below, after briefly presenting the observations of
the beryllium isotopic ratio Be10/Be9 in the energy range $E_0 \gtrsim 1$~GeV.

\subsection{Measurements of the Be10/Be9 ratio for $E_0 \gtrsim 1$~GeV} 
\begin{figure}[t]
\begin{center}
\includegraphics[width=12.7cm]{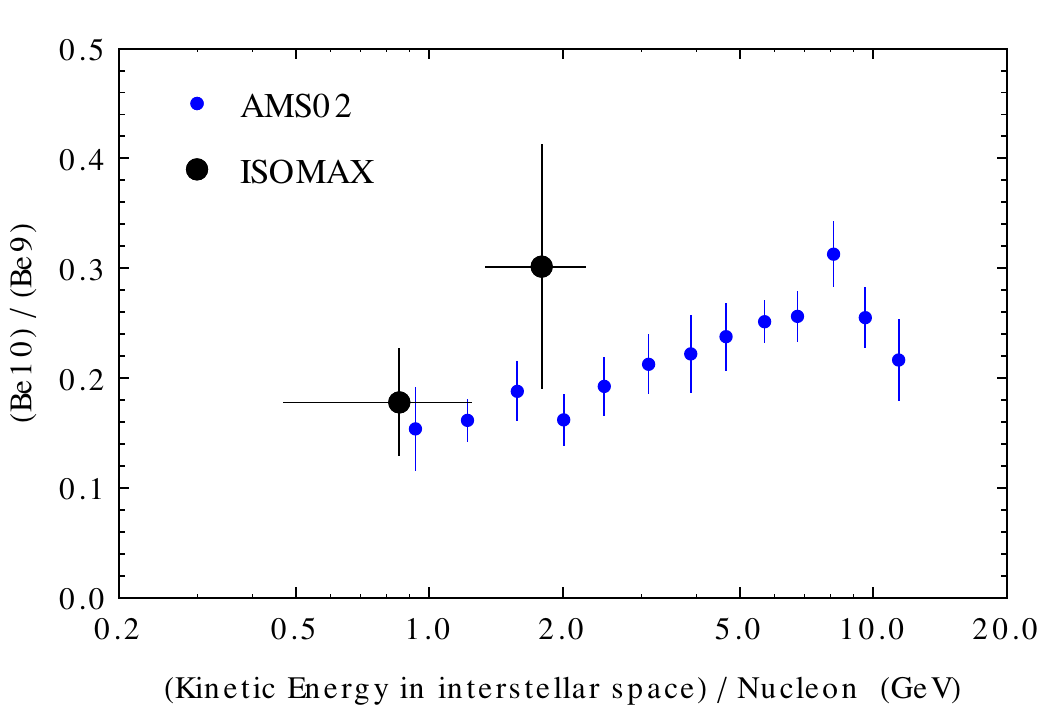}
\end{center}
\caption {\footnotesize
\label{fig:isotopic_ratio} 
Measurements of the isotopic ratio beryllium--10/beryllium--9
at high energy, plotted as a function
of kinetic energy per nucleon. The data is from ISOMAX
\cite{Hams:2004rz} and (only preliminary) from AMS02 \cite{derome-icrc2021}.
 }
\end{figure}

The AMS02 preliminary data \cite{derome-icrc2021}
on the beryllium isotopic ratio Be10/Be9
are shown in Fig.~\ref{fig:isotopic_ratio}, together with the
data of the ISOMAX balloon experiment \cite{Hams:2004rz}, that has
also published a measurement of this ratio above $E_0 \simeq 1$~GeV.

The AMS02 measurement of the isotopic ratio grows slowly from
$R \simeq 0.15$ at the lowest energy to $R \simeq 0.32\pm 0.03$ at 
$E_0 \simeq 8$~GeV,
then the ratio for the next two points is smaller,
and at the highest energy ($E_0 \simeq 11$~GeV)
the ration takes the value $R \simeq 0.22 \pm 0.04$.
These results can be well described with a simple logarithmic dependence:
\begin{equation}
 R_{\rm AMS} (E_0) \simeq (0.16 \pm 0.02) + (0.11 \pm 0.03)
 \; \log_{10} \left( \frac{E_0} {\rm GeV} \right ) ~.
\label{eq:R-ams}
\end{equation}
Combining quadratically statistical and systematic errors, this
(purely phenomenological) fit corresponds to
an acceptable $\chi^2_{\rm min} \simeq 8.2$ for 11 d.o.f.
It is however tempting to speculate that the isotopic 
ratio grows more slowly for $E_0$ close to 10~GeV.
In fact eliminating the three highest energy points,
the best fit has $\chi^2_{\rm min} = 1.9$,
with a reduction of 6.3 units.
The existence of such an effect 
has only a weak statistical significance,
but if real, would have important implications.

The ISOMAX data points have large errors, and correspond
to broad energy bins and therefore provide a weaker constraint
on the isotopic ratio. Using again a simple logarithmic form for the energy dependence
of the ratio the data can be (roughly) represented as:
\begin{equation}
 R_{\rm ISO} (E_0) \simeq
 (0.25 \pm 0.06) + (0.20 \pm 0.12)
 \; \log_{10} \left( \frac{E_0} {\rm GeV} \right ) ~,
\label{eq:R-isomax}
\end{equation}
with a best fit that is a little larger than for AMS02 and grows
more rapidly with energy.

It should be noted that a measurement of the beryllium isotopic
composition above 1~GeV has also
been obtained by the superconducting magnet instrument
for light isotopes (SMILI) \cite{Ahlen-beryllium-2000}.
This experiment has found that out of 26 observed beryllium events,
seven are of Be10, and this, according to the authors, corresponds to
a survival probability consistent with unity,
and has been interpreted as an upper limit
on the ``mean lifetime of cosmic rays'' of
6~Myr at 97.5\% confidence level.
We will not discuss further the SMILI results,
that should however be kept in mind.

Other measurements of the beryllium composition have been
obtained at lower energy
\cite{connell-beryllium-1998,yanasak-beryllium}.

\subsection{Solar modulation effects}
\label{sec:solar_modulations}
The measurements of the isotopic ratio are
performed in the vicinity of the Earth, where the CR spectra
are distorted by time dependent solar modulation effects.
It is convenient to correct for these, 
reasonably well understood effects, 
and obtain the isotopic ratio in the local interstellar medium.
To calculate this correction we have used
the so called force field approximation (FFA)
that describes the solar modulations effects 
assuming that (positively charged) CR particles
during propagation from the boundary of the 
heliosphere to the Earth lose an amount of energy 
proportional to their electric charge:
$\Delta E \simeq e \, Z \, V (t)$, where $V(t)$ is a time dependent potential
associated to the heliospheric electromagnetic fields.
The spectra at the Earth and in the local interstellar medium
are then related by the equation:
\begin{equation}
 \phi_\oplus (E) = \frac{p^2} {p_0^2} ~\phi_{\rm LIS } (E + \Delta E)
\label{eq:ffa}
\end{equation}
where $p$ and $p_0$ are the momenta that correspond
to the energies $E$ and $E +\Delta E$.

Information on the CR spectra at the boundary of the heliosphere have been obtained
by the Voyager satellite, and comparing with the spectra measured by AMS02,
one finds that Eq.~(\ref{eq:ffa}) can provide a reasonably accurate
description of the (time averaged) beryllium spectra using
a potential $V \simeq 0.50$~GV, a value that is also consistent
with the spectral distortions suffered by protons and other nuclei.

In the FFA model, the total energy loss
of a nucleus due to solar modulations
depends only on its electric charge $Z$, 
but this implies that the energy loss per nucleon 
of different isotopes are not identical. It is however straightforward
to take this effect into account.
Using Eq.~(\ref{eq:ffa}), the beryllium isotopic ratio
in the local interstellar medium
can be written in terms of the observed spectra as:
\begin{equation}
\frac{\phi_{10}^{\rm (LIS)} (E_0)} {\phi_{9}^{\rm (LIS)} (E_0)}
 \simeq
 \frac{(E_0- \Delta E_{9} + m)^2 - m^2} {(E_0 -\Delta E_{10} + m)^2 - m^2} 
~\frac{\phi_{10} (E_0 - \Delta E_{10})} {\phi_{9} (E_0 - \Delta E_9)} 
\end{equation}
where $m$ is the nucleon mass, and $\Delta E_{9,10}$ are the energies
per nucleon lost by the two beryllium isotopes.
Introducing the average energy loss 
\begin{equation} 
 \langle \Delta E_0 \rangle = \frac{\Delta E_9 + \Delta E _{10}}{2} \simeq
 \frac{19}{45} ~e\, V
\end{equation}
and the difference
\begin{equation}
\delta E_0 = \frac{\Delta E_9 - \Delta E _{10}}{2} \simeq
 \frac{1}{45} ~e\, V
\end{equation}
(with $V$ the effective heliospheric potential for the data
taking period considered)
and expanding in first order in $\delta E_0$ one obtains:
\begin{equation}
 \frac{\phi_{10}^{\rm (LIS)} (E_0+\langle \Delta E_0 \rangle)}
 {\phi_{9}^{\rm (LIS)} (E_0 +\langle \Delta E_0 \rangle)}
\simeq \frac{\phi_{10}^\oplus (E_0)} {\phi_{9}^\oplus (E_0)}
 ~\left \{1 - \left [ \gamma_9(E_0) + 
 \gamma_{10}(E_0) + \frac{4(E_0 + m)}{(E_0+ 2 m)} \right] 
 ~\frac{\delta E_0}{E_0} \right \} ~.
\label{eq:beryllium_modulations}
\end{equation}
where $\gamma_{A} (E_0) = - d\ln \phi_A/d\ln E_0$ is
the spectral index for the beryllium isotope with mass number $A$
at the energy $E_0$.
Equation (\ref{eq:beryllium_modulations})
states that the isotopic ratio in the LIS
at the energy $E_0^\prime$ can be obtained using the ratio observed at the Earth
at the lower energy $E_0 = E_0^\prime -\langle \Delta E_0 \rangle$
and applying a correction factor to take into account for the different
distortions suffered by the two spectra traversing the
heliosphere. The beryllium spectra in the range measured by AMS02
decrease with energy, and therefore $\gamma_{9,10} >0$, and the correction
factor is $\le 1$, reflecting
the fact that for the lighter isotope Be9 the energy loss 
per nucleon in the heliosphere is larger,
and the effects of modulations more important.
For an effective heliospheric potential of order 0.5--0.6~GV 
as indicated by the data the correction factor is of order 0.9
for $E_0 \simeq 0.5$~GeV, growing monotonically with energy and approaching
asymptotically unity (for $E_0 \gtrsim 30$~GeV).
It should be noted that using this correction the isotopic ratio
in the LIS grows with energy more rapidly than what is observed near the Earth.

\subsection{Isotopic ratio at injection}
\label{sec:psurv1}
To estimate the average survival probability from the isotopic ratio,
we will make the simplifying assumption that the energy of the beryllium nuclei remains
approximately constant during propagation.
The average survival probability can then be written as:
\begin{equation}
 P_{\rm surv} (E_0) 
 = \left [\frac{\phi_{10} (E_0)}{ \phi_9 (E_0)} \right ] \times
 \left [\frac{\phi_{9} (E_0)}{ \phi_{10}^{(0)} (E_0)} \right ]
 \simeq
 \left [\frac{\phi_{10} (E_0)}{ \phi_9 (E_0)} \right ] \times
 \left \{ \frac {\langle q_{9} (E_0) \rangle }
 {\langle q_{10} (E_0) \rangle } \times
 \frac {\mathcal{P}_{10} (E_0)} {\mathcal{P}_{9} (E_0)} \right \}~.
\label{eq:ratio1}
\end{equation}
In the first equality the probability is written as the product of the
isotopic ratio (in the LIS), times a factor that takes into account for the
difference in flux between the two isotopes, estimated assuming that
also the Be10 isotope is stable.
This correction factor is written in the second equality
as the product of two sub--factors that take into account for differences in the
injection rate and in the propagation for the two isotopes.

To estimate the correction factor associated with injection,
it is safe to assume that all beryllium isotopes
are generated by the same mechanism,
that is the fragmentation of larger mass nuclei
(mostly carbon and oxygen) in collision with target gas,
therefore the injection rate for nuclei of type $j$ can be written as:
\begin{equation}
 q_{j} (E_0, \vec{x}, t) = 
 ~ 4 \pi \, \beta \, c~
 \sum_{i} n_i (\vec{x}) \; \sum_{k>j} \phi_k (E_0, \vec{x}, t) ~\sigma_{k + i \to j} (E_0)
\label{eq:q-secondary}
\end{equation}
where $n_i$ is the density of target particles of type $i$ in
the medium where the particles are propagating,
$\phi_k$ the flux of CR nuclei of type $k$, and 
$\sigma_{k+i \to j}$ the relevant fragmentation cross section.
In these collisions the energy per nucleon of the projectile nucleus
and of its fragments in the final state are approximately equal,
and this is why it is convenient to study the spectra in terms or
this kinematical variable.

Eq.~(\ref{eq:q-secondary}) implies that the injection rates of different isotopes
have essentially the same space and time distributions, and that their ratio
in good approximation is only determined by nuclear fragmentation cross section.
This follows from the fact that the composition of the target gas
is expected to be approximately the same in the entire Galaxy,
and that the relative abundances of different primary CR fluxes
are expected to be close to what is observed locally,
while the absolute values of the target gas density and of the
primary spectra cancel in the ratio.

Uncertainties on the values of the fragmentation cross sections are
however not negligible. Fig.~\ref{fig:q-beryllium1} shows the ratio
$q_{10} (E_0)/q_9 (E_0)$ calculated using for the primary CR fluxes
the carbon, nitrogen and oxygen spectra measured by AMS02
(deconvolving solar modulations with the FFA approximations),
and two parametrizations of the proton--nucleus fragmentation cross sections.
One parametrization is presented in Evoli et al. in \cite{Evoli:2019wwu},
while the other is taken from the numerical code GALPROP
\cite{Moskalenko:1997gh,Strong:1998pw,Moskalenko:2021grh}.

Both models predict constant cross sections at high energy
($E_0 \gtrsim 10$~GeV), and therefore an approximately
constant ratio of the injection rates,
however for the Evoli et al. model this constant is
of order 0.82, while using the GALPROP cross section the
asymptotic ratio is of order 0.60.
Important to note is also the energy dependence of the ratio.
that for the Evoli et al. model grows slowly but monotonically
from $\simeq 0.53$ at $E_0 =0.5$~GeV to $\simeq 0.81$ for $E_0 \gtrsim 10$~GeV,
while for the GALPROP model the injection ratio grows to a 
a maximum $\simeq 0.66$ for
$E_0 \approx 1.25$~GeV, and then decreases taking the asymptotic value
($\simeq 0.60$) for $E_0 > 5$~GeV.

The second factor in Eq.~(\ref{eq:q-secondary}) takes into account the fact
that (even neglecting decay) the propagations of
different isotopes with the same energy per nucleon $E_0$
are not identical. This difference emerges because the isotopes
have different rigidities, and therefore travel along different
trajectories in a magnetic field, and
they also have different absorption cross sections
for collisions with interstellar gas.
The size of the propagation effects 
is model dependent, but one can estimate that
the ratio $\mathcal{P}_{10}/\mathcal{P}_9$ is 
close to unity (and a in fact a little less than unity).

The magnetic rigidities of isotopes
of the same element with the same energy per nucleon are 
proportional to the mass number $A$, so the rigidity of Be10 nuclei
is higher by a factor 10/9 with respect to Be9, and this
is expected to result in a faster escape from the Galaxy.
The rigidity dependence of this effect
is often modelled as of power law form
($\propto \rho^{-\delta}$) with the exponent $\delta$ of order 0.3--0.5,
so that the ratio of the propagation effects 
is of order 0.95--0.97.
In other models the rigidity dependence of the escape is weaker,
and the factor closer to unity.

The absorption cross section
is larger for the higher mass number isotope,
so that the effect is again to reduce the $\mathcal{P}_{10}/\mathcal{P}_9$
ratio, however the larger mass number $A$
is compensated by the fact that in
the beryllium--10 nucleus the space distribution of the
nucleons is more compact, with an electromagnetic radius
$\sqrt{\langle r^2 \rangle} \simeq 2.355$~fm (2.519~fm for Be9),
the net result is that the difference in cross sections is small
(of order few percent).
In the following we will assume for the propagation correction factor
$\mathcal{P}_{10}/\mathcal{P}_{9}$ a value of unity, and
estimate that this could be an underestimate of only few percent.

Fig.~\ref{fig:psurv_comp} shows an estimate of the average survival probability
obtained from the isotopic ratios measured by AMS02 and ISOMAX
correcting for solar modulations (for the FFA potential $V = 0.5$~GV) and
for the difference in injection rates using the
parametrisation of the fragmentation cross sections of
Evoli et al. \cite{Evoli:2019wwu} and of GALPROP \cite{Moskalenko:2021grh}.
Use of the two models results in some non negligible differences.

\section{The cosmic ray age distribution}
\label{sec:interpretation}
The problem we will consider now is to interpret
a measurement of the average survival probability
in terms of CR propagation parameters.
Naively, one could estimate an average age for the 
cosmic ray particles $\langle t_{\rm age} \rangle$
using the simple expression:
\begin{equation}
 P_{\rm surv} (E_0)
 \simeq
 \exp \left [ - \frac{\langle t_{\rm age} (E_0)\rangle }{T_{\rm dec} (E_0)} \right ] ~,
\label{eq:t-naive}
\end{equation}
(with $T_{\rm dec} (E_0)$ the decay time at the energy considered).
This simple equation will however give in general incorrect results.
The problem is that the CR particles observed near the solar system
have been injected in interstellar space at different points and at
different times, and therefore their age 
(that is the time interval elapsed
between injection and observation)
is expected to have a broad distribution.
The survival probability must then be obtained calculating the integral:
\begin{equation}
 P_{\rm surv} (E_0) = 
 \int_0^\infty dt~f_{\rm age} (t, E_0) ~e^{-t/T_{\rm dec}(E_0)} ~.
\label{eq:psurv0}
\end{equation}
where $f_{\rm age} (t, E_0)$ is the (normalised) age distribution
(calculated neglecting the effect of decay).
It is then manifest that the survival probability
is determined by the {\em shape} of the age distribution and not only
by a single parameter such as the average $\langle t_{\rm age} \rangle$.
The problem is then to construct a model for the age distribution.

\subsection{The ``leaky box model''}
The ``leaky box'' model has been in use for several decades to
describe CR propagation in the Galaxy. Because of its great simplicity the
model is not really adequate to interpret current observations,
however it can be instructive to consider here its predictions
to illustrate some general points.
The model neglects the space dependence of the cosmic ray spectra,
and describes only a single energy spectrum for each particle type.
Assuming that the energy of the CR particles remain constant
during propagation, the stationary solution
spectrum for an unstable particle is obtained solving the equation:
\begin{equation}
 0 = \frac{dN}{dt} = Q -N\; \left [
 \frac{1}{T_{\rm esc}} +
 \frac{1}{T_{\rm int}} +
 \frac{1}{T_{\rm dec}} \right] = Q - \frac{N}{T_{\rm esc}}
 ~(1 + r + \tau)
\label{eq:leaky-box}
\end{equation}
(where we have left implicit the energy dependence)
that is determined by the three characteristic times for
escape, interaction and decay.
Eq.~(\ref{eq:leaky-box}) is constructed assuming that
the age distribution of the particles, neglecting
the effects of interactions,
is a simple exponential with slope $T_{\rm esc}$.
In the last equality of Eq.~(\ref{eq:leaky-box}) 
we have introduced the notations
$r = T_{\rm esc}/T_{\rm int}$ and $\tau = T_{\rm esc}/T_{\rm dec}$.
The three mechanism of escape, interactions and decay all contribute
to the losses of CR particles, and the 
probabilities for a CR particle to escape, interact or decay are
given by: $1/(1+r + \tau)$, 
$r /(1+ r+ \tau)$ and $\tau/(1+ r + \tau)$.

The average survival probability discussed in this paper
can be obtained comparing spectra calculated including
and neglecting decay and is:
\begin{equation}
 P_{\rm surv}^{\rm (l.b.)}
 = \frac{1 + r }{1 + r + \tau}=
 \frac{1+ \tau \, s}{1 + \tau \,(1 + s)}~
\label{eq:psurv_leakybox}
\end{equation} 
where we have also introduced the adimensional parameter: 
$s = r \, \tau = T_{\rm dec}/T_{\rm int}$.
The average survival probability (for fixed values of
$T_{\rm dec}$ and $T_{\rm int}$) decreases monotonically
for increasing $\tau$ (or equivalently $T_{\rm esc}$),
being unity for short escape times, and reaching a finite asymptotic
value for large $\tau$ (or $T_{\rm esc} \to \infty$):
\begin{equation}
 \lim_{T_{\rm esc} \to \infty}
 P_{\rm surv}^{\rm (l.b.)}
=
 \frac{s }{1 + s } = \frac{T_{\rm dec}}{T_{\rm dec} + T_{\rm int}}~.
\end{equation}

Using the leaky box model to interpret the
measurement $P_{\rm surv}(E_0) = P$ 
and estimate the escape time, one obtains an infinite
number of solutions that can be parametrised
with the assumed value for the interaction time $T_{\rm int}$:
\begin{equation}
 T_{\rm esc}
 = T_{\rm dec} ~\frac{1-P}{P - (1-P)\, s}
\end{equation}
and span the interval:
\begin{equation}
T_{\rm dec} ~\left (1/P -1 \right) \le T_{\rm esc} < \infty~.
\label{eq:tesc_leaky}
\end{equation}
The shortest estimate for $T_{\rm esc}$
corresponds to a very long (diverging) interaction time,
that is to propagation in a very low density medium where interactions
are negligible. Longer values of $T_{\rm esc}$ correspond
to shorter interaction times, and a divergent value of $T_{\rm esc}$
corresponds to a shortest possible value of the interaction
time [$T_{\rm int} = T_{\rm dec} \; (1/P -1)$], and therefore
to an upper limit for the density of the CR propagation medium:
\begin{equation}
 n_{\rm ism} \le \frac{1}{(\sigma_{\rm abs} \, \beta \, c \, T_{\rm dec})} ~
 \left (\frac{1}{P} -1 \right )
\label{eq:nism_leaky}
\end{equation}
(with $\sigma_{\rm abs}$ the absorption cross section).

The simple results discussed above,
that a measurement of the average survival
probability correspond to a lower limit for the CR age,
and to an upper limit on the average density of the propagation medium
remain valid also for the much better motivated diffusion models
discussed below.

\subsection{Diffusion Models}
The construction of CR Galactic propagation models 
has been a central problem in the field for more than seven decades,
and several authors have discussed models where
the effects of the interstellar magnetic fields are described
in terms of diffusion. The first diffusion model was
in fact introduced already in 1951 by Giuseppe Cocconi \cite{cocconi-1951},
who developed the ideas introduced by Enrico Fermi, 
who first proposed a Galactic origin for cosmic rays \cite{Fermi:1949ee},
allowing for the possibility 
that the CR particles are not permanently confined by the
Galactic magnetic fields,
but can be lost ``partly destroyed by collisions with
interstellar matter and partly by diffusing out of the Galaxy''.
To model CR escape, Cocconi described the Galaxy as a homogeneous sphere
of radius $R$ with a stationary source at the center,
and propagation as isotropic diffusion with a constant
(and implicitly energy independent)
diffusion coefficient. Using the boundary condition
that the CR density vanishes at the border of the Galaxy,
Cocconi could then compute the
power required to generate the observed CR density at the Earth.

These concepts were developed further by Morrison, Olbert and Rossi
\cite{morrison-olbert-rossi-1954},
who improved on the spherical model of Cocconi
assuming a cylindrical Galactic confinement volume and more
realistic space distributions of the CR sources
calculating stationary solutions of the diffusion equation,
assuming again a homogeneous, isotropic diffusion coefficient,
and the boundary condition that the CR density vanishes
at the outer limits of the confinement volume. 

The framework where magnetic propagation in the Galaxy
is described as diffusion has been later extensively discussed,
with the inclusion of other effects, such as energy losses,
interactions, advection and decay,
in the influential textbook by Ginzburg and Syrovatskii
\cite{Ginzburg-book}.
This description of propagation has incorporated into numerical codes
such as GALPROP \cite{Moskalenko:1997gh,Strong:1998pw,Moskalenko:2021grh},
DRAGON \cite{Evoli:2016xgn,Evoli:2017vim},
USINE \cite{Boudaud:2017pyx} and PICARD \cite{Kissmann:2017ehy}.
There is at present a large body of literature that interprets
the measurements of the CR spectra with models based on diffusion.

\section{The ``Minimal Diffusion Model''}
\label{sec:minimal}
In this paper we will consider the simplest possible version
diffusion model.
In this model we will assume that the energy of the CR particles remain
constant after injection, and propagation is described as isotropic
diffusion with a constant diffusion coefficient $D$ (of arbitrary
energy dependence) in the volume (the ``halo'') between
the two planes $z \pm Z_{\rm halo}$ that act as absorbers.
Particle are injected continuously in space and time from
the ``disk'' volume between the planes $z = \pm Z_{\rm disk}$ with a
constant rate $q$. The disk volume is homogeneously filled with 
gas of number density $n_{\rm disk}$, while the gas density
in the region $|z| > Z_{\rm disk}$ vanishes.
The absorption cross section (averaged over the composition
of the target gas) is $\sigma_{\rm abs}$, and the particles can also
decay with characteristic time $T_{\rm dec}$.
The observation point where the CR spectrum is measured has the
vertical coordinate $z_{\rm obs}$.
The model is therefore defined by the set of eight parameters
$q$, $z_{\rm obs}$, $Z_{\rm disk}$, $n_{\rm disk}$, $\sigma_{\rm abs}$, $T_{\rm dec}$,
$Z_{\rm halo}$ and $D$.
We find convenient to introduce the diffusion time
\begin{equation}
T_{\rm diff} = \frac{Z_{\rm halo}^2}{2 \, D}
\end{equation}
that gives the order of magnitude of the CR Galactic residence time,
and to discuss the results of the model
replacing the diffusion coefficient $D$ with the diffusion time
$T_{\rm diff}$.

Of the eight parameters of the model, the injection rate controls the
absolute normalisation of the observed spectrum and
cancels in the expressions for the survival probability,
Three parameters
($z_{\rm obs}$, $Z_{\rm disk}$ and $n_{\rm disk}$) must be estimated
from astrophysical observations of the structure of the Galaxy.
The solar system is very close to the Galactic plane and
we will assume $z_{\rm obs} = 0$. In the numerical work
performed below we will also use $Z_{\rm disk} \simeq 0.15$~kpc and
$n_{\rm disk} \simeq 1$~cm$^{-3}$
with a composition (following Ferriere \cite{Ferriere:2001rg})
formed
by 0.9~hydrogen, 0.0875~helium and 0.0125~metals.

The absorption cross section can be in principle
measured from laboratory experiments, or calculated
using Glauber models from a knowledge of the $pp$ cross sections.
We performed such calculation and obtain for
Be10 collisions with hydrogen, helium and oxygen nuclei
at kinetic energy $E_0 \simeq 5$~GeV (and including fragmentation
reactions without pion production) cross sections of 234, 529 and 1051~mbarn.
with only a weak energy dependence.
This corresponds to an average cross sections of order 
270~mbarn (with a weak energy dependence)
and an interaction time for propagation in the disk:
\begin{equation}
 T_{\rm int}^{\rm disk} = [\beta \, c \, \sigma_{\rm abs} n_{\rm disk}]^{-1} \simeq 3.92~
\beta \; \left [\frac {1~{\rm cm}^{-3}} {n_{\rm disk}} \right ]
 \left [\frac {270~{\rm mbarn}} {\sigma_{\rm abs}} \right ]~{\rm Myr}
\label{eq:sigabs} 
\end{equation}
The decay time is readily obtained including relativistic effects:
\begin{equation}
T_{\rm dec} (E_0) \simeq 2.001 \; (1 + 1.073~E_0)~{\rm Myr}
\end{equation}
(with $E_0$ the kinetic energy per nucleon in GeV).
One can note that the decay and interaction time can be of comparable
length, ad therefore that it is important to consider carefully
how they contribute to the formation of the CR spectra of
unstable particles.

The free parameters of model that
must be determined from observations of CR properties are then
the diffusion time $T_{\rm diff}$ and the
halo size $Z_{\rm halo}$

The motivation for using this very simple model
is that it captures the main
properties (and possess the main limitations)
of more complicated models, but it is sufficiently simple
to allow to obtain exact analytic solutions for several interesting
quantities, including the average survival probability,
and this can be very valuable to develop an understanding
of the problem.

\subsection{Escape time and age distributions}
In the leaky box model, discussed in the previous section,
the quantity $T_{\rm esc}$ is equal (neglecting interactions)
to both the average escape time from the Galaxy and the
average age, that is time elapsed from
injection to observation for the CR particles.
In more realistic models these two quantities (escape time and age)
do not coincide, and are also not uniquely defined,
because the first one depends
on the injection point, and the second one depends
on the observation point.

\begin{figure}[t]
\begin{center}
\includegraphics[width=12.7cm]{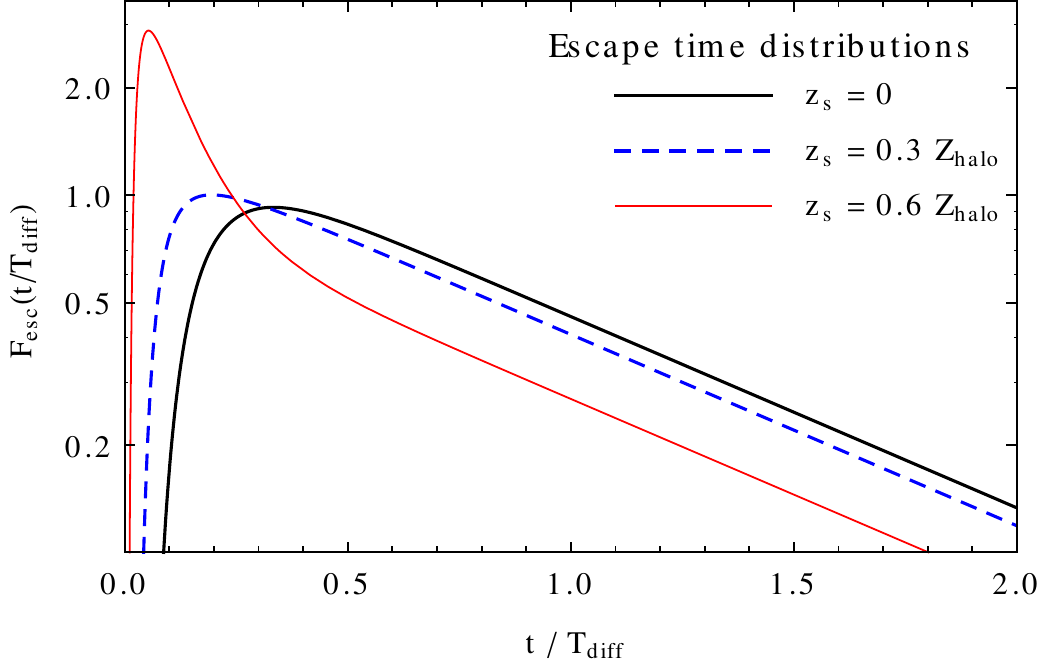}
\end{center}
\caption {\footnotesize
\label{fig:escape} 
Escape time distribution for the simple diffusion model,
shown as a function of the ratio $t/T_{\rm diff}$.
The three curves correspond to 
three different injection points ($z_s/Z_{\rm halo} = 0$, 0.3 and 0.6).
The distribution at large $t$ becomes asymptotically an exponential
with slope $T^*= (8/\pi^2) \, T_{\rm diff}$. 
}
\end{figure}

In the Minimal Diffusion Model, if interactions are neglected,
it is straightforward 
to calculate (in the form of a series) the
distributions for both characteristic times
(see \cite{Lipari:2014zna}).
Some examples of the escape time distributions
are shown in Fig.~\ref{fig:escape}.
The average $\langle t_{\rm esc} \rangle$ can be calculated exactly,
and is:
\begin{equation}
 \langle t_{\rm esc} (z_s) \rangle
 = T_{\rm diff} ~\left (1 - \frac{z_s^2}{Z_{\rm halo}^2} \right )
\end{equation}
(with $z_s$ the vertical coordinate of the injection point),
a result that clarifies the physical meaning associated to
the diffusion time.

\begin{figure}[t]
\begin{center}
\includegraphics[width=12.cm]{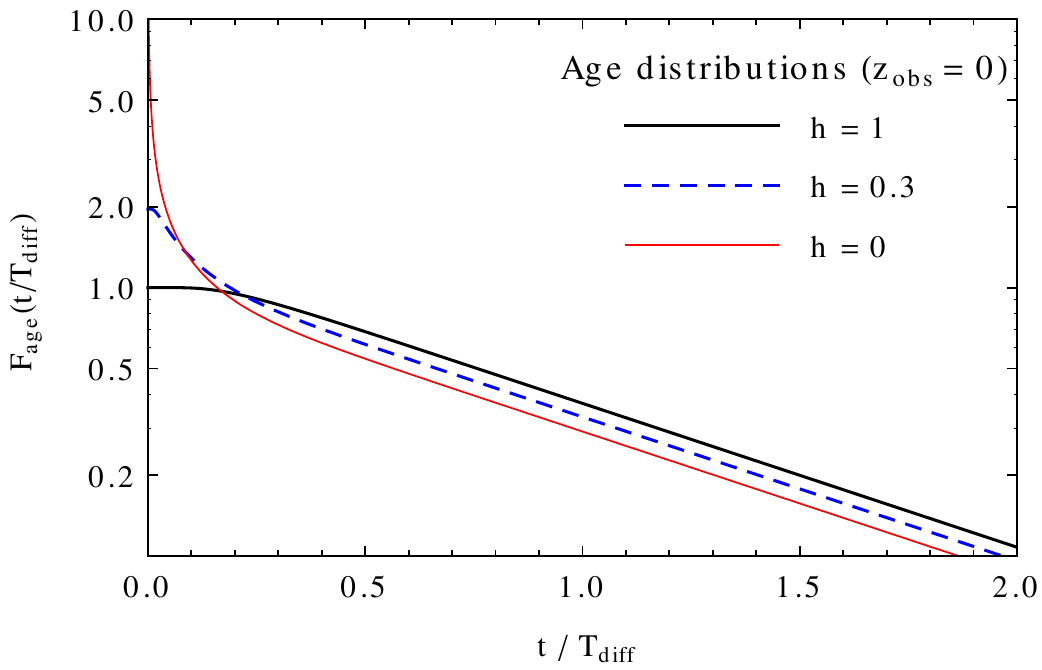}

\vspace{0.15 cm}

\includegraphics[width=12.cm]{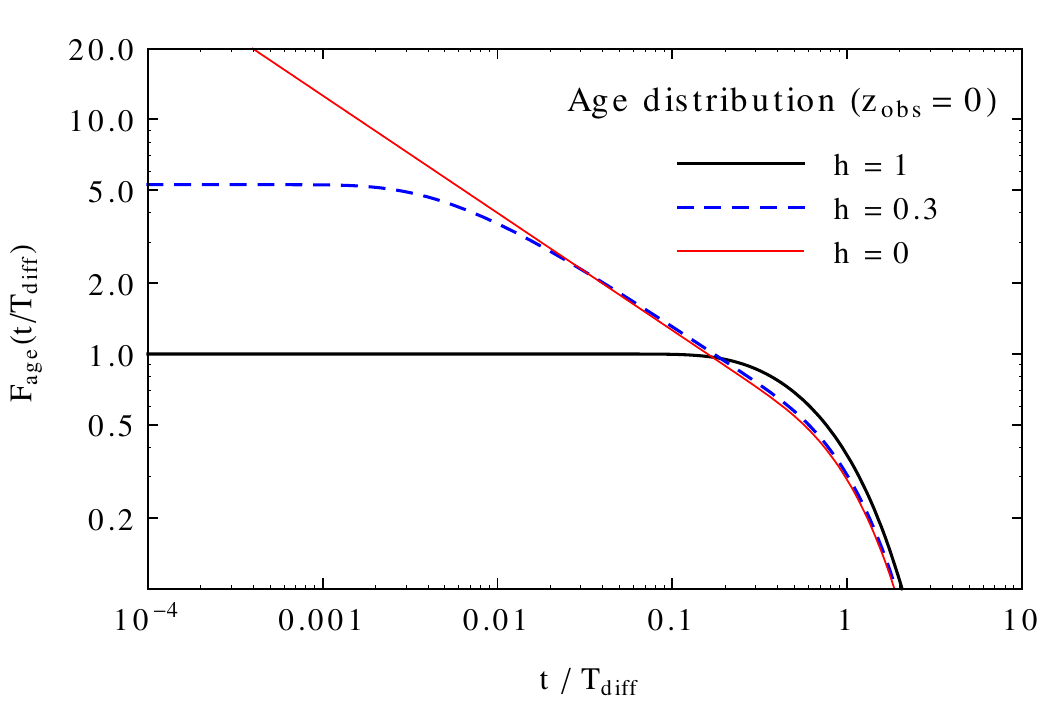}

\end{center}
\caption {\footnotesize
\label{fig:age} 
Top panel: age distribution in the simple diffusion model, shown
as a function of $t/T_{\rm diff}$. The curves are calculated
for an observation point on the Galactic plane ($z_{\rm obs} = 0$)
and for three choices of the ratio $Z_{\rm disk}/Z_{\rm halo}$ (1, 0.3 and 0). 
Bottom panel: the same curves are shown plotted with a log-log scale .
}
\end{figure}

The average age is a function of the observation point, but also
on the space and time distributions of the injection.
For an injection that is constant in time and continuous in space
in the volume $|z| \le Z_{\rm disk}$, the age distribution,
neglecting interactions, 
has been calculated in \cite{Lipari:2014zna}, and some examples 
are shown in Fig.~\ref{fig:age}.
The average age is proportional to $T_{\rm diff}$, and depends on
on the vertical coordinate of the observation point, and on the ratio
$Z_{\rm disk}/Z_{\rm halo}$. For an observation point on the Galactic
disk ($z_{\rm obs} = 0$) one has:
\begin{equation}
 \langle t_{\rm age} (z_{\rm obs} = 0, \sigma_{\rm abs} = 0) \rangle = T_{\rm diff} ~\frac{2}{3} \;
 \left [ 1 + \frac{h}{2} - \frac{h^2}{4} \right ]
\label{eq:tage}
\end{equation}
(with $h = Z_{\rm disk}/Z_{\rm halo}$).
Therefore the average age is of the same order, but not identical to
the average escape time.

Including the effects of interactions the average age
decreases because 
long trajectories suffer more absorption,
and their contribution is suppressed.
A general expression of $\langle t_{\rm age} \rangle$ valid for
an arbitrary observation point and any
value of the ratio $h$ and of the absorption cross section
can be obtained, but is not given here because it is raher complicated. 
The simple expression valid for the interesting case 
of a small $h$ and $z_{\rm obs} = 0$ is:
\begin{equation}
 \langle t_{\rm age} (z_{\rm obs} = 0, h\to 0) \rangle
 = T_{\rm diff} ~\frac{2}{3 + 6 \, r }
\label{eq:tage-int}
\end{equation}
where the adimensional quantity $r$:
\begin{equation}
r = \beta \, c \, T_{\rm diff} \, \sigma_{\rm abs} \, n_{\rm disk} \, h ~
\end{equation}
has the simple physical meaning of the
the average number of interactions during a diffusion time,
calculated after diluting uniformly the interstellar gas
in the entire confinement volume.

An interesting quantity is the average column density
(or grammage) crossed by the CR particles.
Neglecting the effect of interactions,
and for $z_{\rm obs} = 0$, one has:
\begin{equation}
 \langle X \rangle = m \; \beta c \, T_{\rm diff} \; 
 n_{\rm disk} ~\frac{4 \, h} {2 -h}
 \; \left (1 - \frac{7}{6} h + \frac{3}{8} h^2 \right )
 = X_{\rm disk} \; 
 \frac{\beta \, c \, T_{\rm diff}}{Z_{\rm halo}}
\; \frac{2} {2 -h}
\; \left (1 - \frac{7}{6} h + \frac{3}{8} h^2 \right )
\label{eq:xgrammage}
\end{equation}
(where $m$ is the average mass of nuclei in
the interstellar gas, and $X_{\rm disk} = 2 \, Z_{\rm disk} \, n_{\rm disk} \, m$
is the vertical grammage of the disk).
Dividing this grammage by the average 
age given in Eq.~(\ref{eq:tage}) one finds that the observed CR particles
have traveled in a medium of average density 
\begin{equation}
\langle n_{\rm traj} \rangle 
= n_{\rm disk} ~3 \, h \;\frac{2}{(2-h)} \;
\left (\frac{1 - 7/6\, h + 3/8 \, h^2}{1 - 1/2 \, h + 1/4 \, h^2} \right )
 \simeq
n_{\rm disk} ~ 3 \, h \left (1 - \frac{7}{6} h
+ \frac{21}{24} h^2 + \ldots \right )
\label{eq:ntraj}
\end{equation}
This equation states that for small $h$ (that is for
$Z_{\rm halo} \gg Z_{\rm disk}$) the CR particles have traveled
in a medium with an average density that is three times
what is obtained diluting uniformly the gas in the entire halo volume.

It is important to note that Eq.~(\ref{eq:ntraj})
gives a global average, and trajectories of different
length encounter different average densities.
All particles have their origin 
and are observed inside the Galactic disk, 
therefore for very short trajectories the average density
is $\langle n_{\rm ism} \rangle \simeq n_{\rm disk}$,
increasing the length of the trajectory, 
the average density 
decreases monotonically.

The average survival probability depends not only on
the average age of the CR particles, but also on the shape
of the age distribution. This shape, for the case where interactions
are negligible, has already been discussed in \cite{Lipari:2014zna}
and has the scaling form:
\begin{equation}
 f_{\rm age} (t) =
 \frac{1}{T_{\rm diff}}
 ~F_{\rm age} \left ( \frac{t}{T_{\rm diff}}, \frac{Z_{\rm disk}}{Z_{\rm halo}},
 \frac{z_{\rm obs}}{Z_{\rm halo}} \right )
\end{equation}
and depends only on adimensional ratios.

Examples of the distributions, for an observation point with
$z_{\rm obs} = 0$, are shown in Fig.~\ref{fig:age}
for three values of the ratio $h = Z_{\rm disk}/Z_{\rm halo}$.
Inspecting the figure one can easily see the main features of the
age distribution that can be summarised writing:
\begin{equation}
f_{\rm age} (t) \propto
\begin{cases}
{\rm const.}
 & {\rm for}~~ t \lesssim T_{\rm diff} \, h^2 ~, \\[0.12cm]
t^{-1/2}
 & {\rm for}~~ h^2 \lesssim t/T_{\rm diff} \lesssim 1 ~, \\[0.12cm]
e^{-t/T^*}
& {\rm for}~~ t \gtrsim T_{\rm diff}
~~~({\rm with}~~T^*= 8/\pi^2 \, T_{\rm diff}) ~. 
\end{cases}
\end{equation}
One can therefore identify three ranges of $t$ where the
distributions has different forms.
For long times ($t \gtrsim T_{\rm diff}$) the distribution has
an exponential shape, with slope $T^* = 8/\pi^2 \, T_{\rm diff}$.
In the range $h^2 \lesssim t/T_{\rm diff} \lesssim 1$
the distribution grows rapidly for shorter times $\propto t^{-1/2}$.
Finally, for very short times ($t/T_{\rm diff} \lesssim h^2$)
the distribution becomes a constant.

For the limiting cases
$h =1$ and $h \to 0$ the asymptotic forms 
(for short and long times) of the age distributions have simple expressions.
For $h=1$ one has:
\begin{equation}
 F_{\rm age}^{(h= 1)} (x) \simeq
 \begin{cases} 
1 & {\rm for } ~~ x \lesssim 0.5 \\[0.2 cm]
\frac{4}{\pi} ~\exp \left [ - \frac{\pi^2}{8} \, x \right ]
& {\rm for } ~~x \gtrsim 0.5 ~,
 \end{cases}
\label{eq:fage1}
\end{equation}
for $h=0$:
\begin{equation}
 F_{\rm age}^{(h=0)} (x) \simeq
 \begin{cases} 
1/\sqrt{2 \pi \, x} & {\rm for}~~ x\lesssim 0.5 \\[0.2 cm]
\exp\left [- \frac{\pi^2}{8} \, x \right ] & {\rm for}~~ x \gtrsim 0.5 
 \end{cases}
\label{eq:fage0}
\end{equation}
In Eqs.~(\ref{eq:fage1}) and~(\ref{eq:fage0}) (where $x = t/T_{\rm diff}$)
the expressions are exact asymptotically, in the limits of small and large $x$,
but are also a good approximation (better than 10\%)
of the correct result in the entire $x$ range.

\subsection{Average survival probability for a purely magnetic propagation}
It is straightforward to calculate the average survival
probability for the case of a purely magnetic propagation,
that is neglecting the effect of interactions \cite{Lipari:2014zna}.
The result can be expressed in terms of the adimensional parameters
$\tau = T_{\rm diff}/T_{\rm dec}$ and $h = Z_{\rm disk}/Z_{\rm halo}$:
\begin{equation}
 P_{\rm surv}^{\rm no~int.} (\tau, h) =
\frac{1 - \cosh [\sqrt{2 \tau}(1-h)] \;(\cosh [\sqrt{2 \tau}])^{-1}}
{\tau \, h (2 - h)}
\label{eq:psurvdiff} .
\end{equation}
For the limiting cases $h=1$, and $h=0$ one has:
\begin{eqnarray}
 P_{\rm surv} ^{\rm no~int} (\tau, h = 1) & = & 
\frac{1}{\tau} \;
\left ( 1 - \frac{1}{\cosh[\sqrt{2 \tau}]} \right ) \simeq \frac{1}{1 + \tau}
\label{eq:psurv_h1}
\\[0.15 cm]
 P_{\rm surv} ^{\rm no~int} (\tau, h = 0) & = & 
\frac{\tanh [\sqrt{2 \tau}]}{\sqrt{2 \tau}} \simeq \frac{1}{\sqrt{1 + 2 \tau}}
\label{eq:psurv_h0}
\end{eqnarray}
In these equations first equality is an exact result,
while the second (approximate) one 
gives a simpler analytic forms 
that has the correct asymptotic behaviours for large and small $\tau$
and differ from the correct expressions by less that 10\%
in the entire range of definition.
It is elementary to derive the simple expressions for the
survival probability from
the expressions for the age distribution given in
Eqs.~(\ref{eq:fage1}) and~(\ref{eq:fage0}).

One can note that the case
$h = 1$ when the confinement and source volume coincide
the age distribution is essentially indistinguishable from
the ``leaky box'' model, while for small $h$
the age distribution has a large contribution of short times
and for $T_{\rm diff}/T_{\rm dec}$ small the
survival probability is much larger than the leaky box model prediction.

The approximate forms for the average
survival probability given in Eqs.~(\ref{eq:fage1}) and~(\ref{eq:fage0})
can be inverted, so that a measurement $P_{\rm surv} = P$
can translated into a diffusion time
with closed form expressions:
\begin{eqnarray}
T_{\rm diff} (h=1) & = & 
T_{\rm dec} ~\left ( \frac{1}{P} -1 \right )
\\
T_{\rm diff} (h=0) & = & 
T_{\rm dec} ~\frac{1}{2} \; \left ( \frac{1}{P^2} -1 \right )
\end{eqnarray}
These results can be compared to the estimate of the age
obtained with the assumption that the distribution is narrow and
centered at the value $T_{\rm age}$:
\begin{equation}
T_{\rm age} \simeq T_{\rm dec} ~(-\ln P) ~.
\end{equation}
These results (also shown in Fig.~\ref{fig:psurv_models}) illustrate
how the estimate of the age is strongly model dependent.

\begin{figure}[t]
\begin{center}
\includegraphics[width=12.0cm]{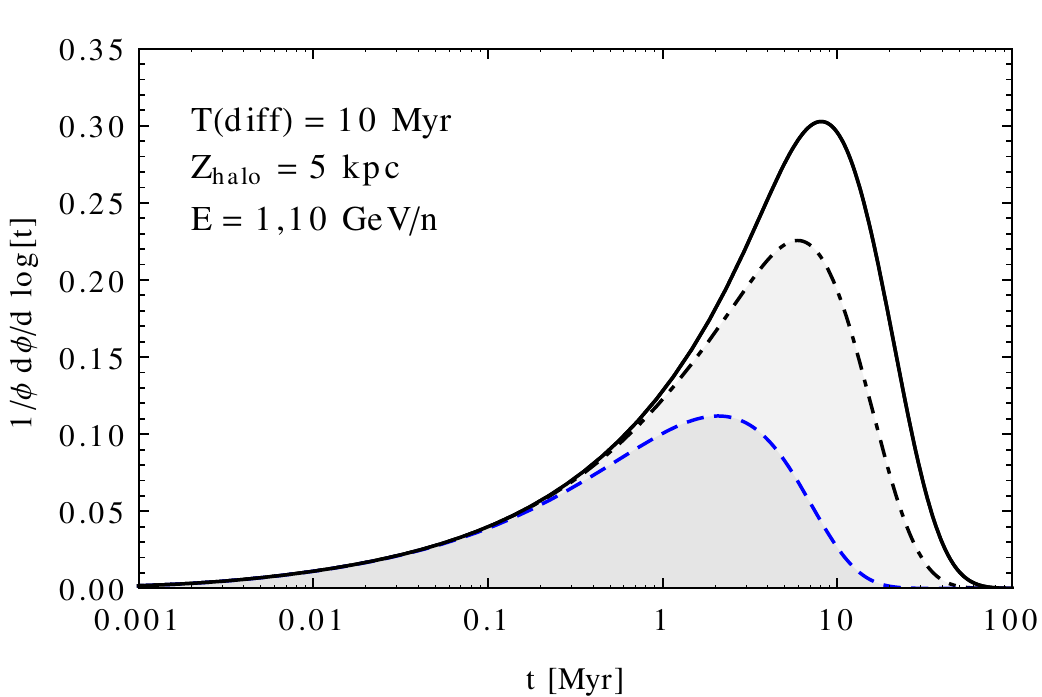}
\end{center}
\caption {\footnotesize
\label{fig:dist_psurv}
The solid line shows an example of the age distribution
calculated in the simple diffusion model
(for the choice of parameters
$T_{\rm diff} = 10$~Myr, $Z_{\rm halo} = 5$~kpc, $s =0$).
The dashed (dot--dashed) line is the distribution calculated
including the effects of decay for Be10 nuclei
with kinetic energy per nucleon $E_0 = 1$~(10)~GeV.
The average survival probability is the ratio between the areas
calculated including and neglecting the effects of decay.
}
\end{figure}

The relation between the age distribution and the average survival
probability is illustrated in Fig.~\ref{fig:dist_psurv} that shows
one example of the age distribution and the effects of decay
on the distribution for two values of $T_{\rm dec}$.
In the figure the value of the survival probability can be easily 
visualised as the ratio of the areas below the curves
that include and neglect decay.

As already stated, the age distribution, and therefore 
the average survival probability, depends on the space and
time distributions of the injection. The results obtained in this
section have been calculated assuming an injection 
continuous in time and space, as it is the case when the main
source of beryllium nuclei is the fragmentation of 
primary cosmic rays in interstellar space.
An alternative possibility
\cite{nested-leaky-box,
Cowsik:2013woa,
Cowsik:2016wso,
Cowsik:2016bwg,
Lipari:2016vqk,
Lipari:2018usj,
Lipari:2019abu}
is that the nuclei are generated by collisions inside or in the vicinity
of CR accelerators.
In this case the injection is not continuous, because the
accelerators are very likely discrete and transient astrophysical
objects, active for only a short time.
In this scenario it is also not possible to
make a unique prediction for the properties of the injection, in part because
the CR accelerators have not been firmly identified,
and also because the sources are of stochastic nature, and one can only
predict their average properties, and the contribution
to the observable spectrum of short age particles, generated by very near
and very young sources can have very large fluctuations.

\begin{figure}[t]
\begin{center}
\includegraphics[width=12.0cm]{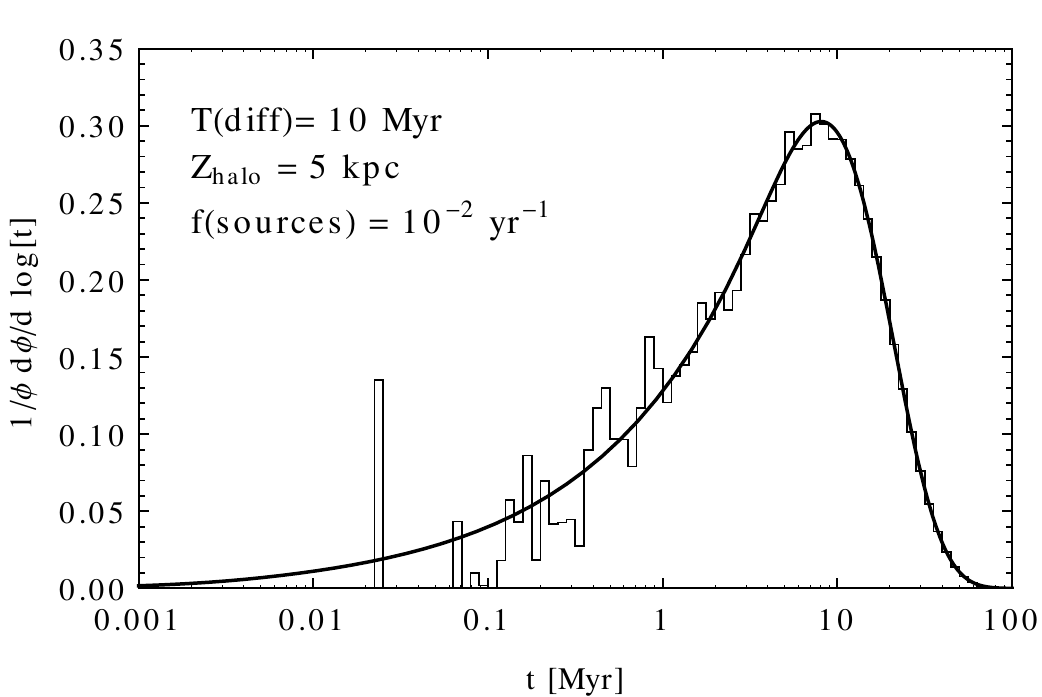}
\end{center}
\caption {\footnotesize
\label{fig:dist_sources}
The solid line shows an example of the age distribution
calculated in the simple diffusion model
(for the choice of parameters
$T_{\rm diff} = 10$~Myr, $Z_{\rm halo} = 5$~kpc, $s =0$), calculated assuming
that the injection of CR particles is continuous in space and time.
The histogram is one Montecarlo realisation of the age distribution 
calculated assuming that the injection has (after averaging) the
same distribution, but is formed by an ensemble
of discrete, instantaneous and point--like events. 
}
\end{figure}

This problem is illustrated in Fig.~\ref{fig:dist_sources}, where
the continuous line is the age distribution calculated for
a continuous (in space and time) injection, while the histogram
is one realisation of a discrete source model generated
with Montecarlo methods assuming that the injection is
formed by a set of discrete, instantaneous emissions with a rate
of one per century in one disk of radius 15~Kpc
(a choice motivated by the properties of Supernova remnants).
This injection model is constructed so that the
{\em average} of the age distributions
obtained from different realisations of the ensemble of sources
is identical to the previous (continuous injection) case.
For the Montecarlo realisation of the sources
shown in the figure the contribution of very short ages
(associated to young near accelerator events) is reduced with respect
to the prediction of a continuous injection, and using
Eq.~(\ref{eq:psurvdiff}) to estimate the diffusion time
from the average survival probability results 
in an overestimate of the correct diffusion time of order 4-5\%.

If the injection of the beryllium nuclei is generated in sources,
their stochastic nature will be a source of systematic uncertainty.
If the distribution of the sources has the same
statistical properties of Supernova explosions,
the approximation of assuming a continuous injection
results in most cases to an underestimate of the diffusion time
of order of few percent, because only in rare cases 
one finds a very young source event in the vicinity of the solar system.

\subsection{Average survival probability including interactions}
The effects of interactions on the average survival probability
can be significant if the CR confinement volume is not too large.
For a stationary continuous injection
one can obtain explicit analytic expressions for the
average survival propbaility that can be
written as a function of the adimensional parameters
$\tau$, $h$ and $s$: 
\begin{equation} 
 s = s_{\rm disk} \, h = \frac{T_{\rm dec}\; h} {T_{\rm int}^{\rm disk}} = 
\beta \, c \, T_{\rm dec} \; n_{\rm disk} \, \sigma_{\rm abs} \,
 \frac{Z_{\rm disk}}{Z_{\rm halo}} ~.
\label{eq:sdef}
\end{equation}
The parameter $s$ has the simple physical meaning of the average number of
interactions during a decay time calculated diluting unformly
the interstellar gas in the entire Galactic volume, 
while $s_{\rm disk} = s/h$ is the same quantity for particles
propagating only in the disk.

For an observation point on the Galactic plane,
the average survival probability, including the effects of interactions
takes the form:
\begin{equation}
P_{\rm surv} (\tau, h, s) =
 \frac{s}{h+s} \; \frac{e^{\sqrt{2 \, h \, (h +s) \, \tau}}-1}
 {e^{\sqrt{2 \, h \, s \, \tau}}-1} ~
 \frac{A_1 \, B_1}{A_2 \, B_2}
 \label{eq:psurv_gen}
\end{equation}
where $A_{1,2}$ and $B_{1,2}$ are:
\begin{eqnarray}
 A_1 & = & e^{\sqrt{2 \tau} h (2 + \sqrt{(h+s)/h})} (\sqrt{h} -\sqrt{h+s})
 - e^{2 \,\sqrt{2 \tau}}\, (\sqrt{h + s} - \sqrt{h})
 \nonumber \\[0.2cm]
 A_2 &=& e^{\sqrt{2 \tau} h (2 + \sqrt{h (h +s)})} (\sqrt{h} +\sqrt{h+s})
 + e^{2 \, h \, \sqrt{2 \tau} } (\sqrt{h + s}+ \sqrt{h}) 
 \nonumber \\[0.2 cm]
 B_1 & = & +\sqrt{(2 \, s\,\tau)/h} \;(1-h) -
 e^{2 \,\sqrt{2\, h \, s \, \tau}} ~
 \left (1+ \sqrt{(2 \, s\,\tau)/h} \; (1-h)\right )-1
 \nonumber \\[0.2 cm]
 B_2 & = & -\sqrt{(2 \, s \, \tau)/h} \; (1-h) -
 e^{\sqrt{2\, h \, s \,\tau}} ~
 \left (1+ \sqrt{(2 \, s\,\tau)/h} \; (1-h)\right )+1
\nonumber
\end{eqnarray}
The limit of the survival probability for a very short diffusion time
is (obviously):
\begin{equation}
 \lim_{\tau \to 0} P_{\rm surv}(\tau,s, h) = 1~, 
\label{eq:psurv_tshort}
\end{equation}
the opposite limit, for very long diffusion times is:
\begin{equation}
 \lim_{\tau \to \infty} P_{\rm surv}(\tau,s, h) = \frac{s}{s + h} =
 \frac{s_{\rm disk}} {s_{\rm disk} + 1} 
\label{eq:psurv_tlong}
\end{equation}
This result can be understood noting that for very slow diffusion
(very long $T_{\rm diff}$) the CR particles remain
always confined in the Galactic disk where they 
decay or interact (with a neglible escape probability).
The average survival probability is then controlled
by the relative importance of decay and interaction in the disk region.

\begin{figure}[t]
\begin{center}
\includegraphics[width=12.7cm]{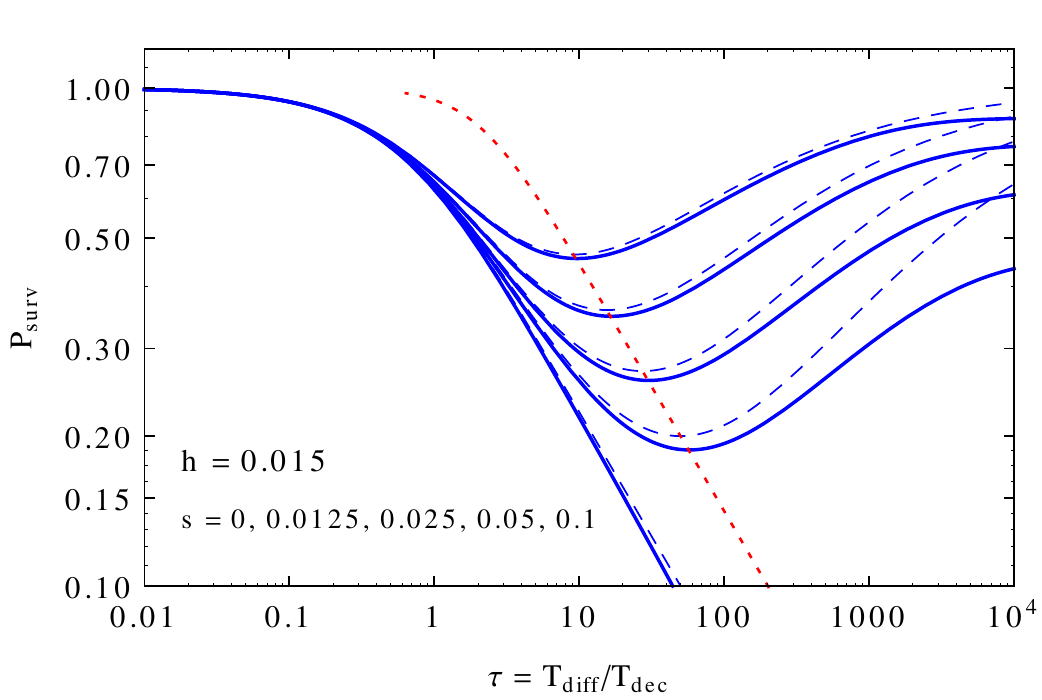}
\end{center}
\caption {\footnotesize
\label{fig:psurv_diff}
Solid lines: average survival probability $P_{\rm surv} (\tau,s, h)$
plotted as function of $\tau$ for one value of the ratio
$h = Z_{\rm disk}/Z_{\rm halo} = 0.015$ and
different values of $s$ ($s = 0$, 0.0125, 0.025, 0.05 and 0.1).
The dashed lines show the average survival probability
for the same values of $s$, but for $h=0$.
The dotted (red) line is the parametric curve
$\{\tau^*(s), P_{\rm surv}^*(s)\}$
[see Eqs.~(\ref{eq:taustar}) and~(\ref{eq:pstar})]
and describes the positions of the minima of the
survival probability for small $h$.
}
\end{figure}

Some examples of the survival probability with the inclusion of the
the effects of interactions are shown in Fig.~\ref{fig:psurv_diff}.
Inspecting the figure one can see that in general
the survival probability has the limits given in
Eqs.~(\ref{eq:psurv_tshort}) and~(\ref{eq:psurv_tlong}),
with one minimum.
The existence of this minimum can be easily understood,
and it corresponds to a diffusion time that is sufficiently long
to have a large decay probability, but not to long, because
the particles must be able to diffuse out of the disk
before being absorbed, and spend time in 
the region of the halo where the target density is low and decay
is favored over interactions.
The minimum decay probability can be well
below the asymptotic value of Eq.~(\ref{eq:psurv_tlong}).

\begin{figure}[t]
\begin{center}
\includegraphics[width=12.7cm]{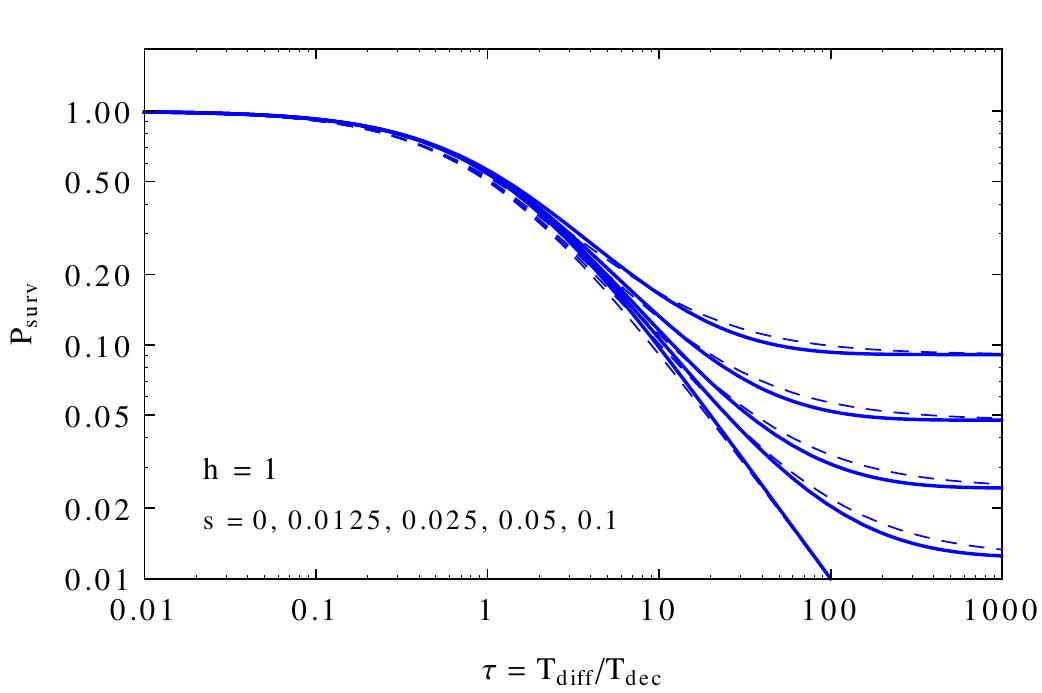}
\end{center}
\caption {\footnotesize
\label{fig:psurv_full}
Solid lines: average survival probability $P_{\rm surv} (\tau,s, h)$
plotted as function of $\tau$ for a constant value $h = 1$ and
different values of $s$ ($s = 0$, 0.0125, 0.025, 0.05 and 0.1).
The dashed lines show the survival probability in the leaky box model.
}
\end{figure}

The Minimum in the survival probability is absent for the 
case $h =1$, when the average survival
probability takes the form:
\begin{equation}
 P_{\rm surv} (\tau,h=1,s) = \frac{s}{s+1} \;
 \frac{(e^{2 \, \sqrt{2 \, s\, \tau}} + 1) \, (e^{\sqrt{2 \, (1+s) \, \tau}} -1)^2}
 {(e^{ \sqrt{2 \, (1+s)\, \tau}} + 1) \, (e^{\sqrt{2 \, s \, \tau}} -1)^2}
\end{equation}
that can be obtained simply substituting the value $h=1$
in Eq.~(\ref{eq:psurv_gen}). In this case the
probability decreases
monotonically from unity at $\tau =0$ to the asymptotic value
$s/(1+s)$ for $\tau = 0$.
This result, shown in Fig.~\ref{fig:psurv_full} is
numerically very close to survival probability for the leaky box model
Eq.~(\ref{eq:psurv_leakybox}).

The distribution in the limit ($h\to 0$)
but for a constant value of the parameter $s$ can be obtained
keeping the halo size fixed, and sending $Z_{\rm disk} \to 0$,
but increasing $n_{\rm disk}$ so that the product
$Z_{\rm disk} \, n_{\rm disk}$ is constant.
The average survival probability then takes the form:
\begin{equation}
P_{\rm surv} (\tau, h=0, s) 
 = \frac{ (e^{2 \, \sqrt{2 \tau}} -1 ) \, (1 + 2 \tau \, s)}
 {(e^{2 \sqrt{2 \tau}}+1 ) \, \sqrt{2 \tau} +
 (e^{2 \, \sqrt{2 \tau}}-1) \, 2 \tau \, s} ~.
\label{eq:psurv_lim0}
\end{equation}
This expression is a good approximation of the average survival
probability for $h$ small, except for $\tau$ much larger than
the value $\tau^*$ where the probability has a minimum.
This is because in the limit $\tau \to \infty$ the expression
in Eq.~(\ref{eq:psurv_lim0}) goes to unity,
in agreement with Eq.~(\ref{eq:psurv_tlong})
for a divergent gas density in the disk.

The interpretation of a measurement of the average survival probability
for Be10 nuclei at kinetic energy per nucleon $E_0$
\begin{equation}
P_{\rm surv} (E_0) = P
\label{eq:interpretation}
\end{equation}
in the framework of the Simple Diffusion Model, is in general 
(without using other considerations) not unique
because there is an infinite number solutions in the form
of pair of values $\{Z_{\rm halo}, T_{\rm diff}\}$, with one quantity
determining the other.
The value of $T_{\rm diff}$ and $Z_{\rm halo}$ that are solutions
of Eq.~(\ref{eq:interpretation}) are in the intervals:
\begin{equation}
\begin{cases}
T_{\rm diff}^{\rm min} \le T_{\rm diff} \le T_{\rm diff}^{\rm max} \\
Z_{\rm halo}^{\rm min} \le Z_{\rm halo} < \infty
\end{cases}
\end{equation}
with the minimum diffusion time $T_{\rm diff}^{\rm min}$
corresponding to a divergent vertical halo size, and the maximum 
$T_{\rm diff}^{\rm max}$ corresponding to the smallest halo size
$Z_{\rm halo}^{\rm min}$.

\begin{figure}[t]
\begin{center}
\includegraphics[width=12.7cm]{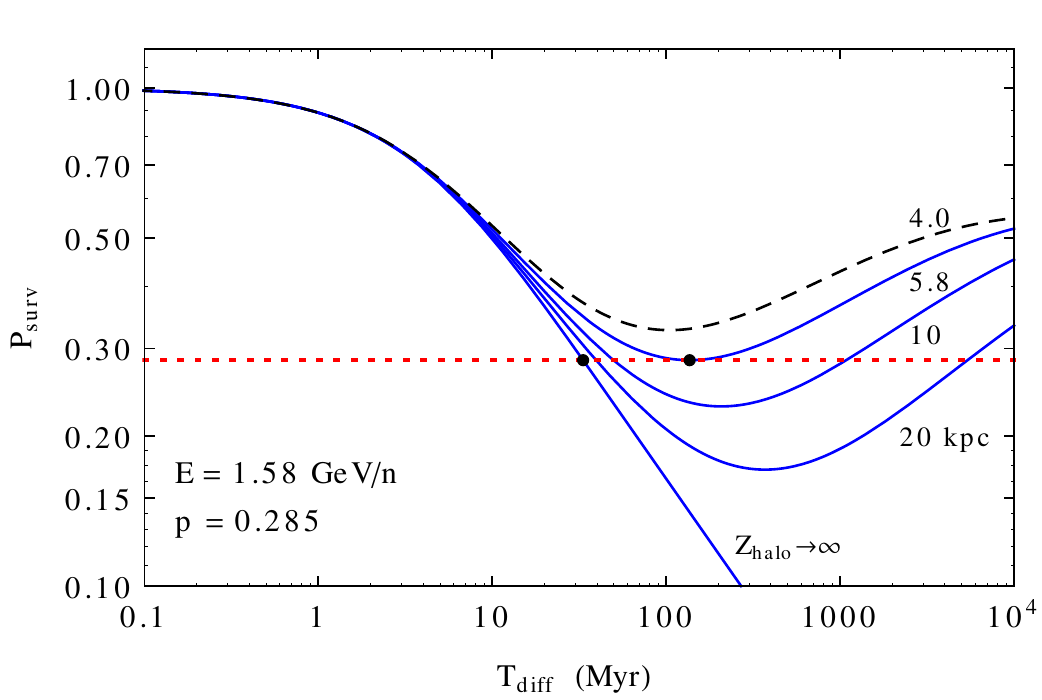}
\end{center}
\caption {\footnotesize
\label{fig:tdiff_estimate}
The solid lines show the average survival probability 
plotted as a function of the diffusion time $T_{\rm diff}$ for
beryllium nuclei of energy $E_0 = 1.57$~GeV/n
(when $T_{\rm dec} \simeq 5.4$~Myr)
for different values of the vertical halo size $Z_{\rm halo}$.
A fixed value of the average survival probability corresponds
to allowed ranges of $T_{\rm diff}$ and $Z_{\rm halo}$.
The figures illustrates the case for
$P_{\rm surv} \simeq 0.285$ (the central value inferred
from the AMS02 measurement using the GALPROP nuclear
fragmentation cross sections).
The minimum allowed $T_{\rm diff}$ corresponds to
$Z_{\rm halo} \to \infty$, while the maximum value
corresponds to the smallest halo size, and also
to the situation where the observed $P_{\rm surv}$ is the
a minimum (for a fixed value of $Z_{\rm halo}$.
}
\end{figure}

An illustration of how a measurement of
$P_{\rm surv}$ corresponds to allowed intervals for
$T_{\rm diff}$ and $Z_{\rm halo}$ is shown in Fig.~\ref{fig:tdiff_estimate}.
The example shown in the figure corresponds to a measurement
$P_{\rm surv} = 0.285$ obtained for $E_0 = 1.57$~GeV/n.
The lower limit for $T_{\rm diff}$ corresponds to
a very large halo size and to a situation where
interactions are negligible. In this case the average survival
probability takes the form of Eq.~(\ref{eq:psurv_h0}), 
and using the approximate form of the probability in
the second equality in Eq.~(\ref{eq:psurv_h0}) one obtains
the simple expression
\begin{equation}
 T_{\rm diff}^{\rm min} (P) \simeq T_{\rm dec}
 ~\frac{1}{2} \left ( \frac{1}{P^2} -1 \right )
\end{equation}
Inspecting Fig.~\ref{fig:tdiff_estimate}, it is easy
to see that the maximum value of the diffusion time $T_{\rm diff}^{\rm max}(P)$
and the corresponding minimum value of the vertical
halo size $Z_{\rm halo}^{\rm min} (P)$
can be calculated from the condition that the survival probability
has a minimum for $T_{\rm diff} = T_{\rm diff}^{\rm max} (P)$.
This corresponds to solving the system of two equations:
\begin{equation}
 \begin{cases}
 P_{\rm surv} \left (
 \tau^* = \frac{T_{\rm diff}^{\rm max}}{T_{\rm dec}} ,
 ~ h^* = \frac{Z_{\rm disk}}{Z_{\rm halo}^{\rm min}},
 ~ s^* = s_{\rm disk} \; \frac{Z_{\rm disk}}{Z_{\rm halo}^{\rm min}} \right ) = P \\
 dP_{\rm surv}/d \tau (\tau^*, h^*, s^*) =0 ~.
 \end{cases}
\label{eq:psurv-system}
\end{equation}

It is straightforward to solve numerically Eq.~(\ref{eq:psurv-system})
to obtain the maximum diffusion time $T_{\rm diff}^*(P)$ and the minimum 
halo size $Z_{\rm halo}^*$ that corresponds to a survival probability $P$.
It is however also possible to have an explicit solution that is a
reasonably good approximation when $Z_{\rm disk}/Z_{\rm halo}^{\rm min}$ is small
(that is in fact the case for the real data).
In this limit the average survival probability can be described 
by the expression in Eq.~(\ref{eq:psurv_lim0}), and the 
position of the minimum ($\tau^*$),
and the value of the probability at the minimum ($P_{\rm surv}^*$)
can be then expressed as a function of the parameter $s$.
The results are reasonably well represented by
the analytic expressions:
\begin{equation}
\tau^* (s) \simeq 2 \; \sqrt{\frac{1}{16 \, (\sqrt{s(s+1)} -s)^4}-1}
\label{eq:taustar}
\end{equation}
and 
\begin{equation}
 P_{\rm surv}^*(s) = P_{\rm surv} (\tau^*, h=0, s)
 \simeq 2 \; \sqrt{\frac{s}{1+ 4 s}}
\label{eq:pstar}
\end{equation}
The line in the plane $\{\tau, P_{\rm surv}\}$
described by the parametric form $\{\tau^*(s), P^*(s)\}$ is
shown as a red dotted line in Fig.~\ref{fig:psurv_diff} and
(to a very good approximation) corresponds to the set of
mimima for the curves $P_{\rm surv} (\tau, h\simeq 0, s)$ for all
values of $s$..
Inverting Eq.~(\ref{eq:pstar}) the parameter $s$ can be expressed
as a function of $P$:
\begin{equation}
 s^*(P) \simeq \frac{P^2}{4 \, (P-1)}
\end{equation}
and this $s$ value can be inserted
in Eqs.~(\ref{eq:taustar}) and~(\ref{eq:sdef})
to obtains explicit expressions for the maximum diffusion time
and minimum halo size:
\begin{equation}
\begin{cases}
 T_{\rm diff}^{\rm max} (P) = T_{\rm dec} ~\tau^*(P) \simeq
 T_{\rm dec} ~2 \; \sqrt{\frac{1}{P^4} -1} ~, \\
 Z_{\rm halo}^{\rm min} (P) \simeq Z_{\rm disk} \;
 [\beta \,c \, T_{\rm dec} \;\sigma_{\rm abs} \, n_{\rm disk}] \;
 \frac{P^2}{4 \, (P-1)} ~.
\end{cases}
\end{equation}

\section{Interpretation of the AMS02 beryllium measurement}
\label{sec:ams}
It can be interesting to study the average survival probabilities
for Be10 nuclei inferred from the preliminaryu AMS02 measurement of the
beryllium isotopic composition in the framework of the Minimal
Diffusion Model discussed in the previous section.

\begin{figure}[t]
\begin{center}
\includegraphics[width=12.7cm]{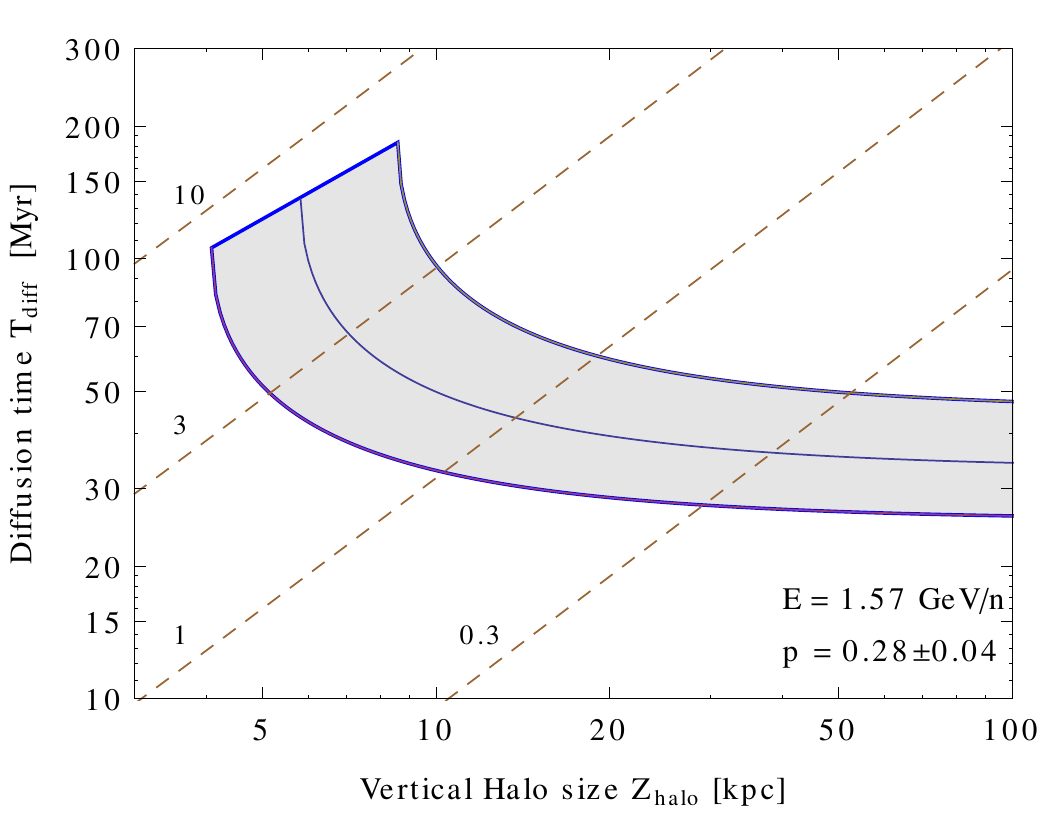}

\vspace{0.12 cm}
\includegraphics[width=12.7cm]{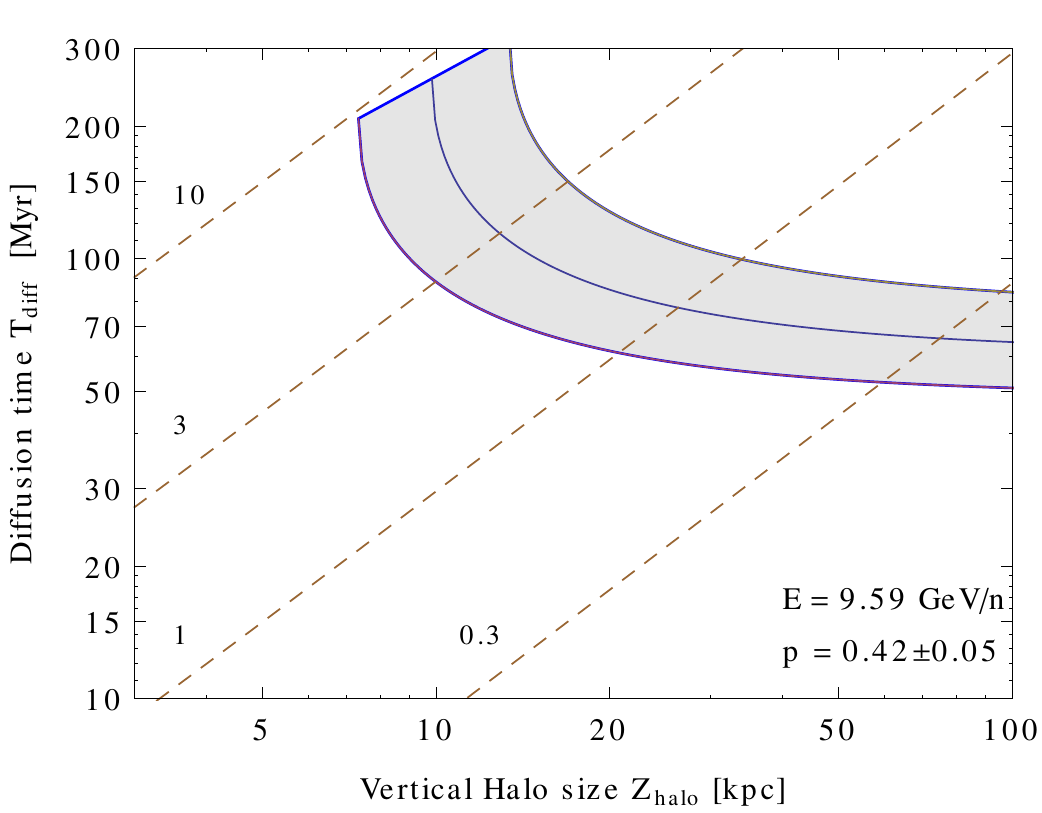}
\end{center}
\caption {\footnotesize
\label{fig:ex1}
Allowed regions in the space $\{Z_{\rm halo},T_{\rm diff}\}$ for a measurement
of the average survival probability in the simple diffusion model.
The top (bottom) panel is for the AMS02 measurement at
$E_0 = 1.57$ and $9.59$~GeV, estimating the average survival
probability with the GALPROP cross sections.
}
\end{figure}

Two examples of this exercise are given in Fig.~\ref{fig:ex1}
that shows (as shaded area) the allowed regions
in the plane $\{Z_{\rm halo}, T_{\rm diff}\}$
obtained finding all solutions of the equation:
\begin{equation}
 P_{\rm surv} \left (
 \tau = \frac{T_{\rm diff}}{T_{\rm dec}(E_0)} ,
 ~ h= \frac{Z_{\rm disk}}{Z_{\rm halo}},
~ s = s_{\rm disk} \; \frac{Z_{\rm disk}}{Z_{\rm halo}} \right )
 = [P(E_0)\pm \Delta P (E_0)]
\end{equation}
where $P(E_0) \pm \Delta P(E_0)$ is the estimate of the average
survival probability obtained from the AMS02 measurement
of the beryllium isotopic composition at kinetic energy per nucleon
$E_0$, and the probability $P_{\rm surv}(\tau,h,s)$ (calculated
in the framework of the Minimal Diffusion Model)
is given by Eq.~(\ref{eq:psurv_gen}).

In Fig.~\ref{fig:ex1} the top (bottom) panel shows
the allowed region for the estimate at
$E_0 \simeq 1.57$~GeV ($E_0 \simeq 9.59$~GeV)
using the fragmentation cross sections in the GALPROP code.
As discussed in the previous section,
the measurement of the survival probability
corresponds to an allowed interval of $T_{\rm diff}$,
with a lowest value that requires
a very large halo size, and a maximum value that
corresponds to the smallest possible halo.

The pair of parameters $\{Z_{\rm halo}, T_{\rm diff}\}$ determines
the average grammage $\langle X \rangle$ traversed by secondary particles,
[see Eq.~(\ref{eq:xgrammage})], and lines of constant
$\langle X \rangle$ are also shown in the figure.

\begin{figure}[t]
\begin{center}
\includegraphics[width=12.7cm]{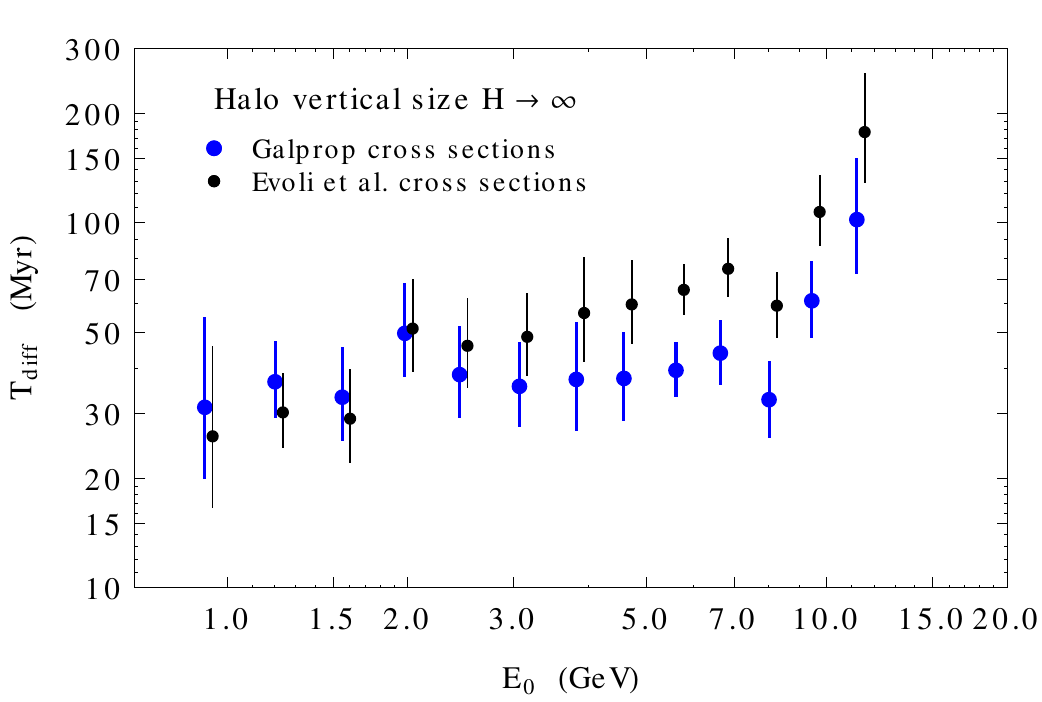}

\vspace{0.12 cm}
\includegraphics[width=12.7cm]{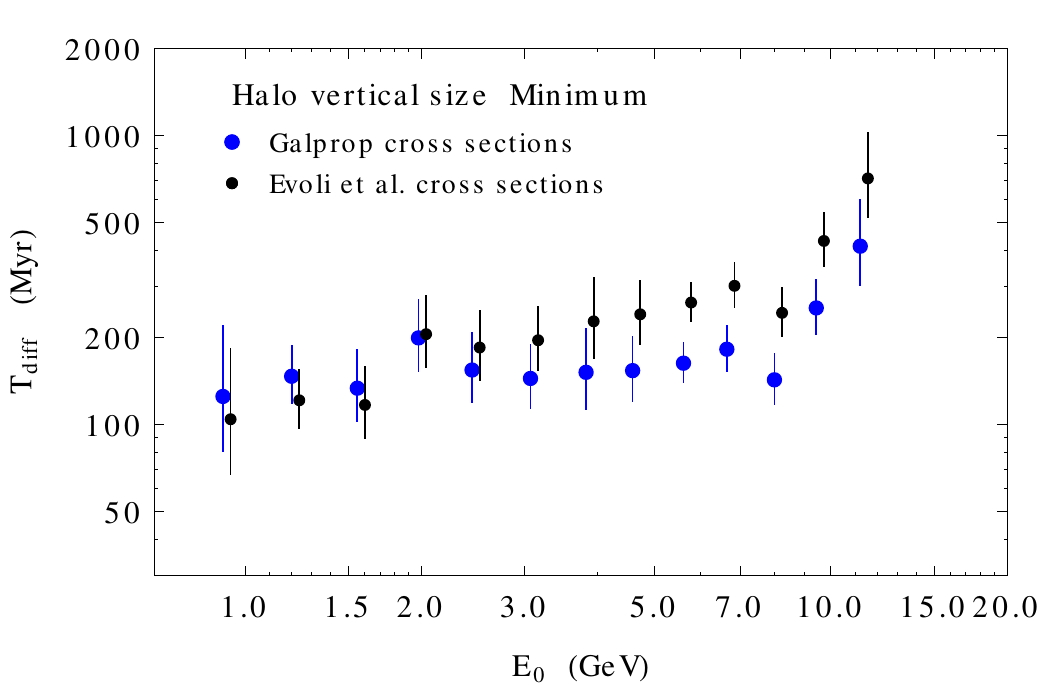}
\end{center}
\caption {\footnotesize
\label{fig:tdiff1}
Estimates of the diffusion times obtained from the AMS02
measurements of the beryllium isotopic ratio in the simple
diffusion model. The survival probability is calculated using the
Evoli et al. \cite{Evoli:2019wwu}, or the GALPROP \cite{Moskalenko:2021grh}
cross sections.
In the top panel the diffusion time is calculated neglecting the
effect of interactions (that is in the limit $Z_{\rm halo} \to \infty$).
In the bottom panel the halo size has the minimum allowed value.
}
\end{figure}

Fig.~\ref{fig:tdiff1} shows a summary of the diffusion times
estimated from the AMS02 measurements interpreted in the
framework of the Minimal Diffusion Model.
The top panel shows estimates of the allowed interval
of $T_{\rm diff}$ calculated assuming a very large halo size,
while the bottom panel shows the
$T_{\rm diff}$ interval calculated for the smallest halo size consistent
with the measurement.
In both cases, for each energy the interval is calculated twice
using the fragmentation cross sections
of Evoli et al. \cite{Evoli:2019wwu} and of GALPROP \cite{Moskalenko:2021grh}.
For both cross section models, if one assumes a
large confinement volume the diffusion time at $E_0 \simeq 1$~GeV
is of order 30~Myr. Increasing the energy ,
for the GALPROP cross sections the diffusion time remains
approximately constant, except for the two highest energy
points ($E_0 \simeq 9.5$--11.5~GeV) where the estimate becomes
a factor 2--3 higher.
Using the Evoli et al. cross sections, that estimate a larger
Be10 fraction at production, the effects of decay must be
larger, and the diffusion time longer, so that
$T_{\rm diff}$ is estimated of order 50--60~Myr when $E_0$ is a few GeV.
The diffusion times estimated for the two highest energy points is
large ($T_{\rm diff} \gtrsim 100$~Myr).

\begin{figure}[t]
\begin{center}
\includegraphics[width=12.7cm]{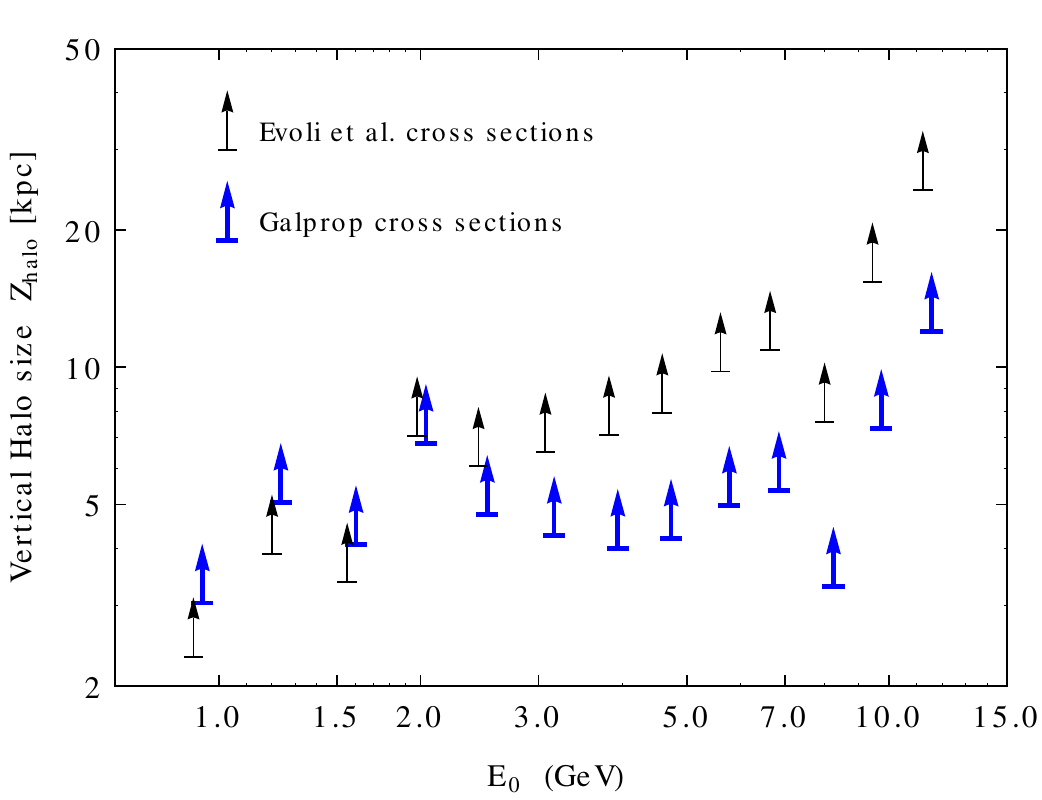}
\end{center}
\caption {\footnotesize
\label{fig:halo_limits}
Lower limit on the CR vertical halo size obtained from the
AMS02 measurement of the beryllium isotopic ratio calculated
in the framework of the simple diffusion model.
The two estimates are obtained using the nuclear fragmentation
cross sections of Evoli et al. \cite{Evoli:2019wwu},
and of GALPROP \cite{Moskalenko:2021grh}.
}
\end{figure}

Fig.~\ref{fig:halo_limits} shows lower limits
on the vertical halo size obtained with the two sets of
fragmentation cross sections for different values of
the energy. The limits are of order of 5~Kpc for
$E_0 \approx 1$--2~GeV, and grow with increasing energy,
being more stringent for the Evoli et al. cross sections.

\section{Outlook}
\label{sec:outlook}
Several works on Galactic cosmic rays
have estimated the properties of their propagation in the Milky Way
from the study of the ratios of the spectra of
secondary (Li, Be and B) and primary (C, O, $\ldots$) nuclei.
This ratio can be interpreted in terms
of the grammage traversed by the CR particles.
For example the HEAO--3 team \cite{Engelmann:1990zz}
used a leaky box model framework to estimate a
rigidity dependent grammage:
$\langle X \rangle \simeq 14.0 \, \beta \, (\mathcal{R}/\mathcal{R}_0)^{-0.60}$
for rigidities $\mathcal{R} > \mathcal{R}_0 = 4.4$~GV
(and a constant value for $\mathcal{R} < \mathcal{R}_0$)
More recently Evoli, Aloisio and Blasi \cite{Evoli:2019iih} 
interpreted the AMS02 data in a diffusion model,
obtaining results very close to those of the HEAO--3 collaboration
($\langle X \rangle \simeq 8.4$~g/cm$^2$ at $\mathcal{R} > 10$~GV,
and a rigidity dependence $\propto \mathcal{R}^{-0.63}$).
If one makes the assumption that the grammage is 
integrated during propagation in interstellar space,
one can then infer the CR Galactic residence time,
or more in general the product of the residence time
and the average density of the interstellar medium along the CR trajectories.
Several authors have recently discussed estimates of
these quantities in the framework of diffusive models
\cite{Boschini:2018baj,
Boschini:2019gow,
Weinrich:2020cmw,
Weinrich:2020ftb,
Genolini:2021doh,
Luque:2021nxb,
Korsmeier:2021brc} obtaining results that are in reasonable
(if not perfect) agreeement with each other.

A very attractive and often discussed idea
is to combine the studies of the secondary/primary ratio
and of the beryllium isotopic
composition to solve the ambiguity in the interpretation of the data
between confinement time and average density, and more in general
to test the assumption that the grammage
is integrated in interstellar space.

We will postpone a detailed discussion of such a combined study
waiting for the publication of the beryllium isotopic composition measurements
by the AMS collaboration.
We can however note that the preliminary study
performed here indicates that there is significant tension
between the standard interpretation of the secondary/primary
ratio and the estimates of the cosmic ray age
inferred from the preliminary AMS02 data.
The diffusion times calculated using
the ``Minimal Diffusion Model'' appear to {\em increase}
with energy, by a factor of order two to three, when the
kinetic energy per nucleon grows from 1 to 10~GeV,
in contrast to the results obtained from
the primary/secondary ratio, that suggest that the diffusion time
should decrease by a factor larger than two in the same energy interval.

The indications of a discrepancy are stronger (weaker)
when the effects of decay are calculated using
the nuclear fragmentation cross sections of Evoli et al.
\cite{Evoli:2019wwu}, (GALPROP \cite{Moskalenko:2021grh}),
and emerge especially from the two highest energy points of the AMS02
measurements.
It is therefore possible that the apparent conflict 
between data and model is the result of 
incorrect estimates of the relevant nuclear cross sections,
and/or of systematic errors in the preliminary data.
It can however be interesting to speculate on the implications
of the case where the results are confirmed,
and the estimates of the fragmentation cross sections remain valid.

It should be noted that the possibility of a conflict between
the data on the beryllium isotopic composition
and the estimates of the cosmic ray age inferred
by the secondary/primary ratios can be understood qualitatively
with simple considerations.
In the AMS02 the isotopic ratio Be10/Be9 grows slowly 
from $R \simeq 0.16$--0.17 for an observed energy
$E_0$ of order 0.7--1.0~GeV
(that corresponds to a energy of order 0.9--1.2~GeV outside the heliosphere),
to $R \simeq 0.25$~for $E_0$ of order 4--6~GeV.
The three highest energy points
(at $E_0 \simeq 8$, 9.5 and 11.4~GeV) have values
$R \simeq (0.31\pm 0.03)$, ($0.25 \pm 0.03$) and ($0.21 \pm 0.04$).
In this energy range the decay time
$T_{\rm dec}$ grows by a factor larger
than six, from 4.1~Myr at $E_0 = 1$~GeV to 25.6~Myr at 11~GeV,
and therefore one expect that even if the CR age
is energy independent, the size the effects of decay
on the unstable Be10 isotope should decrease significantly.
Moreover, the data on the secondary/primary ratio
(in the standard intepretation) suggest
that the CR age decrease as a power law with rigidity
($\langle t_{\rm age} \rangle \propto \beta \, \mathcal{R}^{-\delta}$).
The energy interval of the AMS02 data corresponds to 
the rigidity range $\mathcal{R} \simeq 4.2$--29.7~GV,
resulting in an expected shortening of the CR average age
by a factor that goes from  1.7 (for $\delta = 0.33$) to 2.3
(for $\delta = 0.5$) in the interval of the AMS02 measurements.
The two effects, the longer decay time and the expected shorter
residence time, both go in the direction of reducing the effects of decay,
and therefore one expects the average survival probability
to have a significant increase in the energy range of the
AMS02 measurements.

\begin{figure}[t]
\begin{center}
\includegraphics[width=12.7cm]{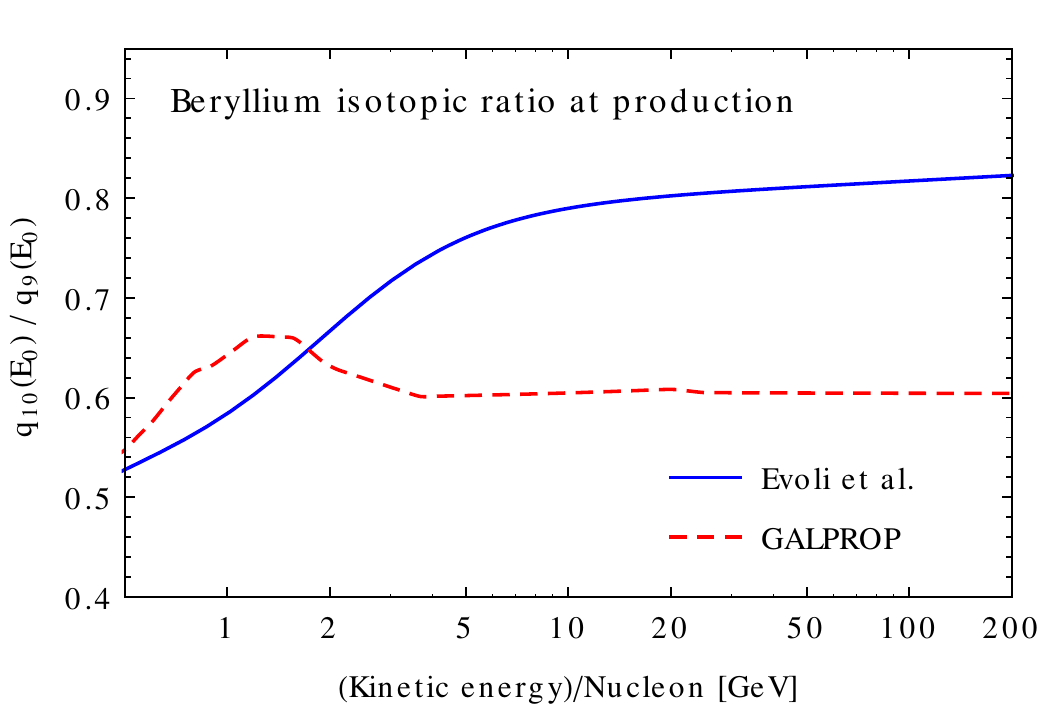}
\end{center}
\caption {\footnotesize
\label{fig:q-beryllium1} 
Ratio of the production rates of beryllium--10 and beryllium--9
plotted as a function of kinetic energy per nucleon.
The ratio is calculated assuming for the interacting cosmic ray particles
the energy spectra measured by AMS02 \cite{AMS:2021nhj}, 
and for the
nuclear fragmentation cross sections the values
tabulated in Evoli et al. \cite{Evoli:2019wwu} and
those in the GALPROP code \cite{Moskalenko:2021grh}.
The calculation includes only the leading contribution
of interactions with an hydrogen target.
}
\end{figure}

The estimate of the average survival probability from the isotopic ratio
depends on the nuclear fragmentation cross section, and therefore
the slow energy dependence of the isotopic ratio could in principle
be the result of a cancellation, with a
Be10/Be9 ratio at production that grows with energy,
and energy decay effects that are stronger at low energy.
For the models of the fragmentation cross sections used in this work
the isotopic ratio at production (shown in Fig.~\ref{fig:q-beryllium1})
has a value of order 0.6--0.8, with only a weak energy dependence.
Assuming the validity of these cross section models
implies that: (i) the effects of decay are significant and suppress
the spectrum of the unstable isotope Be10,
and (ii) the effects of decay have a weak energy dependence,
even when the decay time changes by a large factor.

\begin{figure}[t]
\begin{center}
\includegraphics[width=12.7cm]{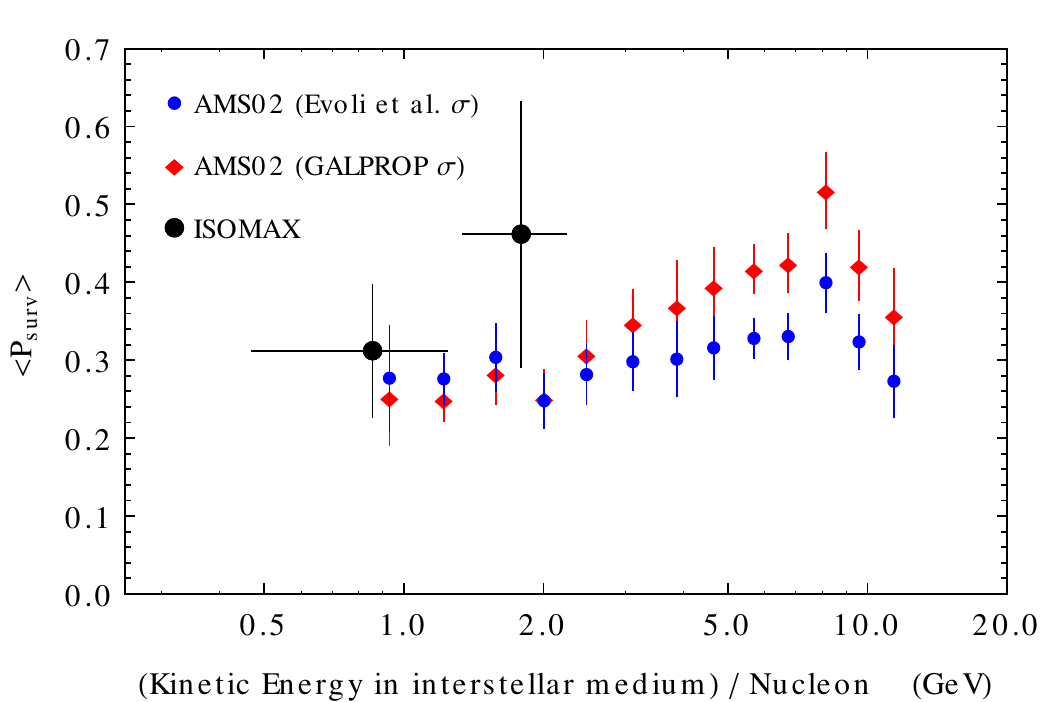}
\end{center}
\caption {\footnotesize
 \label{fig:psurv_comp}
 Estimate of the average survival probability
 $P_{\rm surv} (E_0)$ of beryllium--10 nuclei,
 as a function of kinetic energy per nucleon
 in the local interstellar medium. The probability is estimated
 from the measurements of the beryllium isotopic ratio Be190/Be9 
 by ISOMAX \cite{Hams:2004rz} and AMS02 \cite{derome-icrc2021},
 including corrections for
 solar modulations, and assuming constant energy during propagation
 in interstellar space. The isotopic ratio at injection
 is calculated using the nuclear fragmentation cross sections of
 Evoli et. al. \cite{Evoli:2019wwu} and of GALPROP \cite{Moskalenko:2021grh}.
 }
\end{figure}

The predictions of the two models for
the nuclear fragmentation cross sections have some significant differences.
Using the Evoli et al. cross section model \cite{Evoli:2019wwu}
the isotopic ratio at prodiuction grows
from a value $R_0 \simeq 0.59$ at
for $E_0 \simeq 1$~GeV to an asymptotic value $R_0 \simeq 0.81$ at
asymptotic value of order 0.80 at high energy.
Using the GALPROP model \cite{Moskalenko:2021grh}
the prediction for the isotopic ratio at production 
is $R_0 \simeq 0.64$ at $E_0 \simeq 1$~GeV and decreases to 
an asymptotic value of order 0.60 at high energy, therefore
in this case one obtains an average survival probability
that is a little larger and that grows a little more rapidly with energy.
However, for both models, the growth of $P_{\rm surv}$ is only slow,
and interpreting the results in the framework of
the Minimal Diffusion Model one obtains a
diffusion time that increases with energy (see Fig.~\ref{fig:psurv_comp}).

\begin{figure}[t]
\begin{center}
\includegraphics[width=12.7cm]{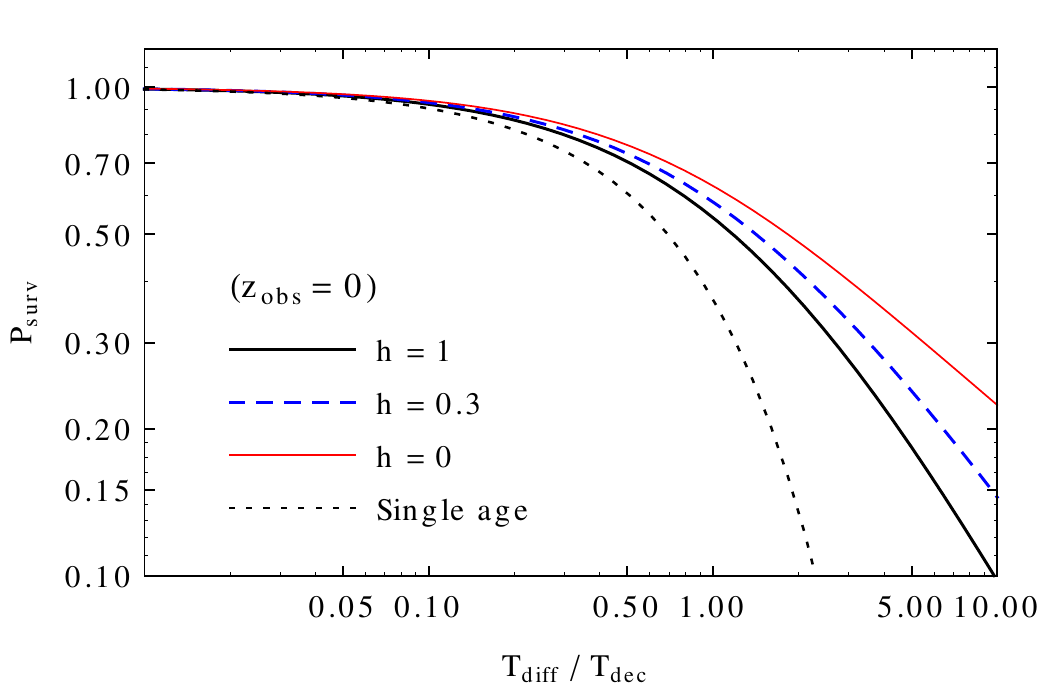}
\end{center}

\caption {\footnotesize
\label{fig:psurv_models} 
Average survival probability calculated for the Minimal Diffusion
Model, for an observation point on the Galactic plane
($z_{\rm obs} =0$), and plotted as a function of the
ratio $T_{\rm diff}/T_{\rm dec}$. The different curves are
calculated for three values of the ratio $h = Z_{\rm disk}/Z_{\rm halo}$
($h = 0$, 0.3 and 1). The dotted line corresponds to a narrow age distribution.
 }
\end{figure}

An average cosmic ray age that increases with energy
is not only in conflict with the standard interpretation
of the secondary/primary ratio that estimate a
grammage that decreases with rigidity, but is also very difficult
to understand constructing a model for the magnetic structure
of the Galaxy.
The alternative is to modify the theoretical framework for
the interpretation of the results on $P_{\rm surv} (E_0)$.
In this work have shown (see for example Fig.~\ref{fig:psurv_models})
that the same value of $P_{\rm surv}$
can correspond to very different values of the average age
of the CR particles in different propagation models.
Similarly, a measurement of the energy dependence of
the survival probability $P_{\rm surv} (E_0)$ can be interpreted
with different energy dependences of the propagation parameters
in different propagation models.

It is easy to see that a survival probability that changes very
slowly with energy can be consistent with an average age 
that is constant  or  change very slowly with energy
if the shape of the age distribution is very broad.

If the CR age distribution is broad, 
the average survival probability $P_{\rm surv}$
takes (in first approximation) the physical
meaning of the fraction of the observed particles
with age in the interval $t_{\rm age} \lesssim T_{\rm dec} (E_0)$.
The decay time grows linearly with the Lorentz factor of the nuclei,
and therefore, for a constant shape of the age distribution,
$P_{\rm surv}$ increases with energy because
the time interval where decay is important becomes smaller.
This growth of $P_{\rm surv}$ with energy
is slower for a broader distribution.

In the Minimal Diffusion Model the age distribution is
determined by two parameters the diffusion times
and the halo vertical size.
It is however possible for the age distribution to have
a more complicated shape that depends on more parameters (that could have
different energy dependences).
The preliminary AMS02 data (interpreted with current
models of the fragmentation cross sections)
indicate that when the decay time grows from approximately 4~Myr to
approximately 30~Myr the average survival probability
remains in rather narrow range 
($P_{\rm surv} \simeq 0.25$--0.4) suggesting a very
broad age distribution where large fractions of particles have 
ages that are both very short ($t_{\rm age} \lesssim$~few~Myr)
and very long ($t_{\rm age} \gtrsim 50$~few~Myr).
This broad age distribution could exist
if the CR confinement volume
is formed by an inner halo and a more extended halo (perhaps
associated with the existence of the Fermi bubbles) that have
confinement times of different orders of magnitude.

The estimate of the CR age distribution is crucially
important for the interpretation of the electron
and positron spectra, in particular to establish the
existence of a new source of relativistic positrons
\cite{Lipari:2016vqk,Lipari:2018usj,Lipari:2019abu}.
A sufficiently long CR age implies that the large rate of energy losses
for $e^\mp$ spectra will result in a strong softening
of their spectra, and therefore that the observed hard positron
spectrum cannot be generated by the secondary production mechanism
and requires a harder source.
The preliminary AMS02 beryllium data, as interpreted in the
previous section, indicate a CR age that seems to be
in conflict with the hypothesis 
of secondary production for CR positrons.
This conclusion is again based on the validity of
the current estimates of the nuclear fragmentation cross sections.
An approximately constant isotopic ratio for beryllium
could in principle be consistent with energy
independent fragmentation cross sections and with
a short CR age, so that the decay effects for Be10
are small in the entire energy range considered.
This interpretation however requires that the observed
isotopic composition is equal to the one generated at injection,
and this hypothesis is at present strongly disfavoured.

The modeling of nuclear fragmentation cross sections
is the main source of systematic uncertainties
in extracting the very valuable information
encoded in the beryllium isotopic composition.
Reducing these uncertainties with an appropriate program
of experimental studies is very desirable and of great value.

\vspace{0.35 cm}
\noindent {\bf Acknowledgments}\\
I'm grateful to Pedro De la Torre Luque for help in obtaining the GALPROP
nuclear fragmentation cross sections, and to
Carmelo Evoli and Michael Korsmeier for interesting discussions.

\clearpage


\clearpage

\clearpage

\end{document}